\documentclass[11pt,a4paper]{article}
\usepackage{float}
\usepackage{jheppub}

\usepackage[T1]{fontenc}
\usepackage[utf8]{inputenc}
\usepackage{microtype}
\graphicspath{{figures/}}

\usepackage{amsthm,mathtools}
\usepackage{bm}
\usepackage{xcolor}
\usepackage{enumitem}
\usepackage{array}
\usepackage{tikz}
\usepackage{booktabs}
\usetikzlibrary{arrows.meta,positioning}

\hypersetup{
  pdftitle={Collider event distances from the phase space sphere},
  pdfauthor={Nathaniel Craig, Brian Henning, Andrew J. Larkoski,
             and I\~nigo Valenzuela Lombera}
}

\allowdisplaybreaks[1]
\definecolor{darkblue}{rgb}{0,0,0.5}
\definecolor{pastelpurp}{RGB}{140,84,208}

\newtheorem{theorem}{Theorem}[section]
\newtheorem{proposition}[theorem]{Proposition}
\newtheorem{lemma}[theorem]{Lemma}
\newtheorem{corollary}[theorem]{Corollary}
\theoremstyle{definition}

\newtheorem{remark}[theorem]{Remark}
\newtheorem{example}[theorem]{Example}

\newcommand{\R}{\mathbb{R}}
\newcommand{\norm}[1]{\left| #1 \right|}
\newcommand{\abs}[1]{\left| #1 \right|}
\newcommand{\ip}[2]{\left\langle #1,#2\right\rangle}
\newcommand{\cC}{\mathcal{C}}
\newcommand{\cK}{\mathcal{K}}
\newcommand{\dd}{\,\mathrm d}

\DeclareRobustCommand{\okina}{%
  \raisebox{\dimexpr\fontcharht\font`A-\height}{%
    \scalebox{0.8}{`}%
  }%
}

\title{Collider event distances from the phase space sphere}

\author[a,b]{Nathaniel Craig,}
\author[c]{Brian Henning,}
\author[d]{Andrew J.\ Larkoski,}
\author[a,1]{and I\~nigo Valenzuela Lombera%
\note{Corresponding author.}}

\affiliation[a]{Department of Physics,
University of California, Santa Barbara,
CA 93106, USA}

\affiliation[b]{Kavli Institute for Theoretical Physics,
University of California, Santa Barbara,
CA 93106, USA}

\affiliation[c]{Department of Physics and Astronomy,
University of Hawai\okina i, Honolulu,
HI 96822, USA}

\affiliation[d]{American Physical Society,
Hauppauge, New York 11788, USA}

\emailAdd{ncraig@ucsb.edu}
\emailAdd{bhenning@hawaii.edu}
\emailAdd{larkoski@aps.org}
\emailAdd{inigo@physics.ucsb.edu}

\abstract{
Can the distribution of collider events in phase space be used to
distinguish the processes that produced them? We present a distance-based classification method that uses the global
topology of Lorentz-invariant phase space and retains reconstructed masses.
To tackle particularly challenging classification tasks that may benefit from higher event resolution, we develop a polynomial-time approximation to compute the distances. The resulting classifiers achieve strong discrimination across several
simulated benchmarks, including fully hadronic four-top production
against $t\bar t+\text{jets}$.
}
\begin{document}
\maketitle
\flushbottom

\section{Introduction}

Collider observables are functions on phase space that probe the dynamics of the underlying field theory, making phase space itself an essential substrate for understanding the Standard Model and searching for physics beyond it.  A textbook understanding of phase space focuses almost solely on its volume measure, as that is all that is needed to calculate cross sections and decay rates in quantum field theory~\cite{Peskin:1995ev,Srednicki2007,Schwartz:2014sze,ParticleDataGroup:2026mpi}.  In this paper, we take a step back and use the global topology of relativistic phase space as a starting point for constructing event distances. Treating phase space geometrically enables a sensible definition of distances between entire collider events. Progress towards the understanding and application of phase space has recently been made in Refs.~\cite{LM2020, HenningMelia}. We find significantly improved performance on classification tasks relative to previous applications of phase space distances~\cite{CCKL2024}.

\begin{figure}[H]
  \centering
  \small
  \includegraphics[width=\linewidth]{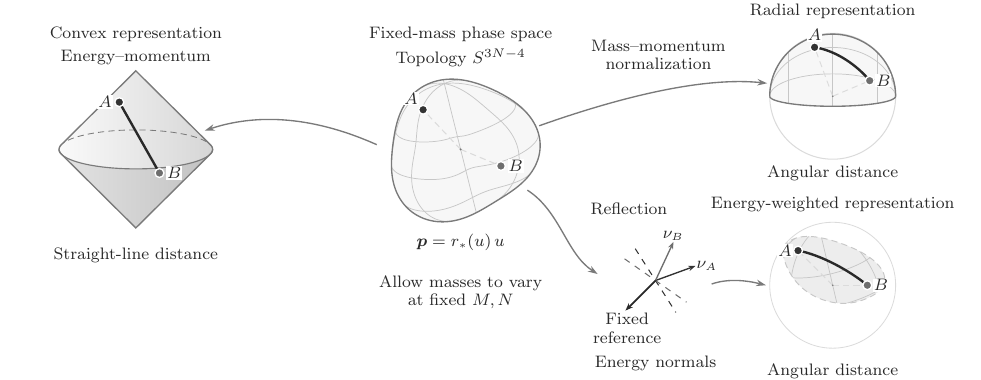}
  \caption{Schematic construction of event distances from phase-space geometry. For $M > \sum m_i$, where $M$ is center-of-mass energy, fixed-mass phase space has spherical topology. Momentum is decomposed into polar coordinates $\bm{p} = r_*(\bm u)\bm u$. Allowing reconstructed masses $m_i$ to vary leads to a convex representation and several distinct spherical representations of collider events.}
  \label{fig:event-geometries}
\end{figure}

Lorentz-invariant phase space ($\Phi_N$) is the domain of integration for the matrix element in a relativistic Quantum Field Theory. Cross sections and other observables are expressed as integrals over phase space,
\begin{equation*}
    \sigma\propto\int_{\Phi_N} |\mathcal{M}(\bm p)|^2 \,d\Phi_N(\bm p)
\end{equation*}
Much of the attention naturally goes to the matrix element, and correctly so, as it encodes all the dynamics of the theory. Phase space enforces the kinematic constraints of spacetime and the particle mass shells, and it isn't precisely \textit{ignored} in the broader literature: it is central in event generation and Monte Carlo integration, unitarity, EFT bases, amplitude bases and many other areas~\cite{RAMBO, MADGRAPH,LM2020, HenningMelia2019Harmonics, shu2022constructinggeneralpartialwaves,SrivastavaSudarshan1958,BogoradEtAl2026}. For collider applications,  phase-space geometry has also informed anomaly detection, through topology-aware latent spaces~\cite{NgairangbamEtAl2025} and event-level observables for weakly supervised searches~\cite{Brennan:2025WeaklySupervised}. More recently, relative representations constructed from massless phase-space and Wasserstein distances have been combined with particle-identification information for binary tagging of individual boosted jets~\cite{CaiLi2026M3R}. For event classification purposes, the matrix element tells us the distribution of events in this high dimensional space. Therefore, a natural question is: \textit{does the event distribution in phase space carry information useful for classification?}

Ref.~\cite{CCKL2024} studied this exact question in massless phase space  by using lightcone coordinates on an open domain~\cite{LM2020}, which follow from a spinor-frame description~\cite{HenningMelia}. That classifier has already been
used in a CMS search for microscopic black holes and electroweak sphalerons,
including BlackMax signals against QCD
multijets~\cite{VamiZhang2025,CMSBlackHoles2026}. 
Alongside other recent distance-based approaches (including the energy mover's
distance~\cite{KMT2019}, linearized and unbalanced optimal transport~\cite{LOT2020,WhichMetric2022}, spectral metrics~\cite{Spectral2023}, and multiscale constructions for complete events~\cite{Multiscale2025}), distances based on the geometry of phase space complement conventional observables for event classification and anomaly detection~\cite{MLHEPReview,JetReview}. The massless approximation used by the
earlier phase-space classifier~\cite{CCKL2024} discards clustered-object masses and becomes
less accurate at coarse resolution, motivating a comparison
that retains them throughout the resolution scan.

We start from the known spherical topology
of fixed-mass phase space~\cite{Sokolov1971, BatsonEtAl2021}.
We present a simple radial map that assigns each momentum configuration to a unit
vector, defining an angular distance. Including masses among the
coordinates extends this construction to events with different
reconstructed masses. The resulting \textit{radial} representation, its convex \textit{energy--momentum} counterpart, and two \textit{energy-weighted} spherical representations provide the four comparisons studied here, illustrated in Fig.~\ref{fig:event-geometries}. After removing each event's overall
invariant-mass scale, these representations retain the same four-vector
kinematics but weight differences in energy sharing, momenta, and masses
differently. Their classification performance can therefore differ even
though they preserve the same normalized event information.

The resolved multiplicity of a collider event can greatly exceed that of
the hard scattering: showering and hadronization populate additional
momenta, while exclusive clustering resolves the final state into a chosen
number $N$ of pseudojets. Fully hadronic four-top production is a demanding
example: the top-pair-plus-jets background already contains real tops and
substantial radiation, so a top tag alone cannot distinguish the two
processes~\cite{CMSFourTopHadronic2023}.
Semiclassical black-hole signals provide a complementary case, motivating
comparisons of many-particle kinematics without assuming a fixed
resonance-decay chain~\cite{CMSBlackHoles2026}.
Additional radiation can itself help distinguish signal from background:
an ATLAS search for R-parity-violating gluino decays gains sensitivity by
requiring a jet beyond the six-quark decay topology~\cite{ATLASRPVMultijet2024}.
Finer exclusive-$k_T$ resolutions can expose more of this decay and
radiation structure, while at coarser resolutions some of it is retained
in pseudojet masses. The comparison should therefore accommodate the
masses produced by clustering and remain computationally tractable as the
resolved multiplicity increases.

At high multiplicity, explicit minimization over all particle permutations
is computationally prohibitive. We use the \textit{Hungarian
algorithm}~\cite{Kuhn1955} to efficiently find the one-to-one matching of
particles between two events that minimizes a chosen matching cost with
their orientations held fixed, followed by alignment in the transverse
plane. This sequential procedure approximates the joint minimum over
assignment and alignment. We use pairwise distances to train a support vector machine (SVM) on labeled signal
and background events. The same four-vector preprocessing and event
comparisons can be used to train classifiers for different signal and
background processes, without constructing process-specific
discriminating observables.

We study how the event representation and resolved multiplicity affect
discrimination in three particle-level benchmarks: $t\bar t$ versus QCD
dijets, fully hadronic four-top versus $t\bar t+\text{jets}$, and conditioned
semiclassical black-hole signals versus QCD multijets. In the massless $t\bar t$--QCD
comparison, the updated prescription using Hungarian assignment
substantially reduces the earlier method's loss of discrimination at high
multiplicity. Among the complete pipelines compared, the energy-weighted
spherical representation that treats all particles symmetrically achieves
the highest discriminating performance at the largest tested multiplicity in all three benchmarks. Its
discrimination improves or approaches a plateau as the resolution
increases, with four-top and black hole reaching broad plateaus near $N=15$
and $N=20$, respectively. Rankings at lower multiplicity and the useful
resolution remain process dependent.

Section~\ref{sec:fixed-shell} develops the physical event space from
energy and momentum conservation, and variable reconstructed masses, establishing the
basis for spherical and convex comparisons.
Section~\ref{sec:collider-metrics} defines the four mass-aware
representations and their associated metrics.
Section~\ref{sec:implementation} specifies the normalization,
assignment--alignment procedure, and supervised classifier, together with
the massless benchmark and common evaluation protocol.
Section~\ref{sec:results} assesses the resulting discrimination as a
function of resolved multiplicity. The appendices collect numerical and computational details, phase-space measure treatment, formal topology and metricity proofs, and spinor-helicity details.
\section{From phase space to reconstructed events}
\label{sec:fixed-shell}

Our goal is to compare collider events through their structure in phase space. We first establish its topology: its global shape, including how it is connected and whether it has holes or boundaries. We begin with fixed particle masses, then enlarge the space to accommodate the varying masses of reconstructed objects. We also describe the same events in different coordinates, providing complementary views of this enlarged space. This common topological foundation leaves freedom in how differences between events are measured, motivating the distances introduced in Section~\ref{sec:collider-metrics}. Particle matching and alignment are addressed in Section~\ref{sec:implementation}.

\subsection{Fixed-mass phase space}
\label{subsec:fixed-mass-shell}

We can define phase space as the \textit{set} of on-shell momenta that satisfy energy and momentum conservation rather than in the traditional fashion through the phase-space measure. In the center-of-mass (COM) frame, with total momentum $P=(M,\bm0)$, for $N\geq2$ particles, $m_i\geq0$, and $M-\sum_i m_i>0$, define $N$-particle fixed-mass phase space as
\begin{equation}
 \Phi_N(M;\bm m)=
 \left\{\bm p = (\bm p_1,\ldots,\bm p_N)\in(\R^3)^N:
 \sum_i\bm p_i=0,\quad
 \sum_i\sqrt{m_i^2+\norm{\bm p_i}^2}=M\right\}
 \label{eq:phase-space-body}
\end{equation}
where $\bm p_i$ are the momentum 3-vectors. Equation~\eqref{eq:phase-space-body} includes $\bm p_i=0$ when $m_i=0$, and the particle masses are fixed. The motivation for proposing this change in perspective is that correct parametrizations of an integral may discard essential topological information. Removing a measure-zero region like $x=\frac{1}{2}$ from the integration domain in $\int_0^1x \,dx$ does not change its value, but it has topological consequences. The phase-space measure and the treatment of zero momentum are specified in Appendix~\ref{app:fixed-lips}.

The global topology of phase space is particularly simple and well known \cite{Sokolov1971, BatsonEtAl2021}. Early event generation literature used several of its features,\footnote{A particular example of this is massive RAMBO rescaling and event unweighting~\cite{RAMBO}.} even when it isn't directly stated. Topological equivalence (denoted $\cong$) is defined by whether two spaces are \textit{homeomorphic}, meaning we can construct a bijective map from a space to another that behaves continuously in both directions. This map is called a \textit{homeomorphism}. There are other equivalent ways to show topological equality. In this section, we will construct an explicit homeomorphism from a unit sphere onto phase space. Essentially, we are about to show that a ping-pong ball and a balloon are topologically equivalent, but for a higher dimensional shell. Further, a homeomorphism is useful to define a distance on the target that holds on the original space.

\begin{figure}[H]
  \centering
  \small
  \includegraphics[width=0.9\linewidth]{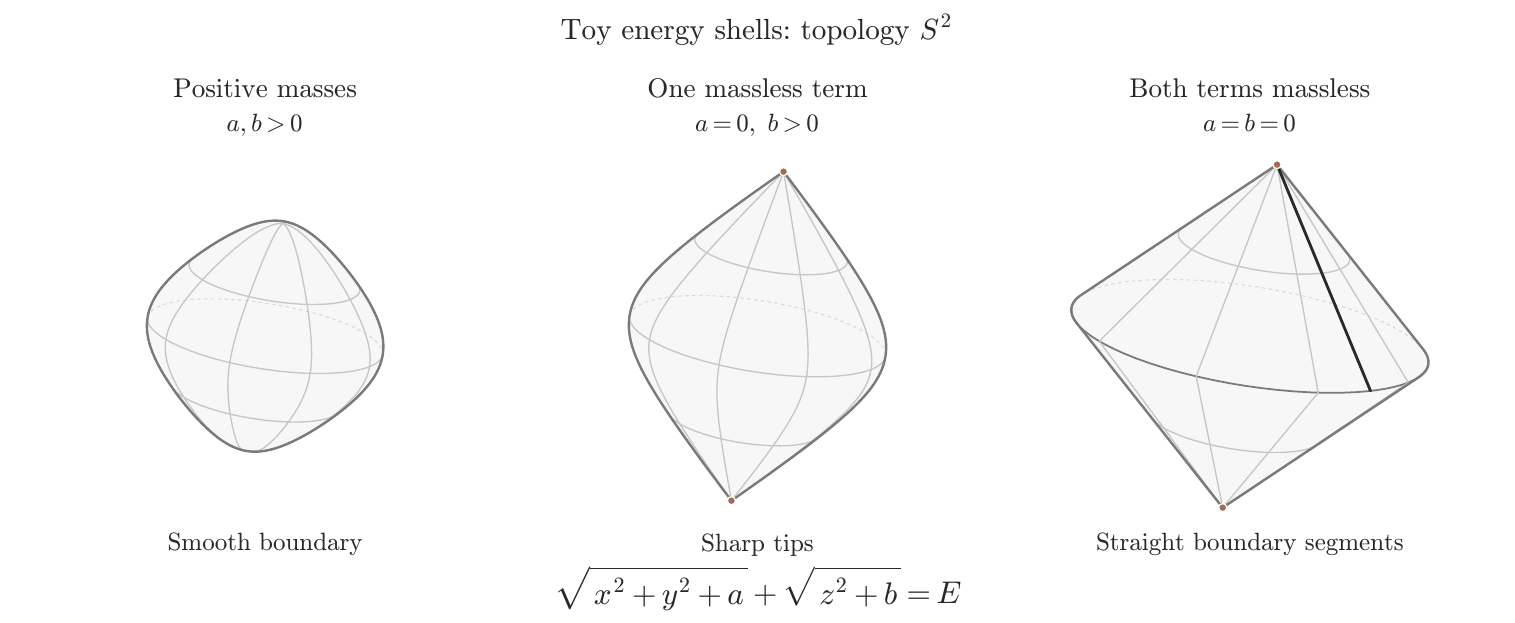}
  \caption{Schematic illustration of fixed-mass phase-space manifold geometry. A shell with massless particles can develop sharp features.}
  \label{fig:sharp-geometries}
\end{figure}

Let $\bm{u}_i = \bm p_i /\norm{\bm p}$ where $\bm p = (\bm p_1, \ldots,\bm p_N)$. These $\bm u_i$ are the normalized momentum components with respect to the \textit{concatenated} vector magnitude, meaning ${\bm p} = r(\bm u)\bm u$ is a polar decomposition, where the radius is in principle allowed to vary with the angular direction. Therefore, $\bm u$ lives on a unit sphere in the ($3N-3$)-dimensional momentum conserving vector space  \begin{equation}
     \hat S=\left\{\bm u=(\bm u_1,\ldots,\bm u_N)\in(\R^3)^N:
 \sum_i\bm u_i=0,\quad\sum_i\norm{\bm u_i}^2=1\right\}
\end{equation}
This is a $(3N-4)$-dimensional unit sphere. Let
$r_*(\bm u)>0$ be the unique solution of the energy conservation constraint at each $\bm u$,

\begin{equation}
\sum_i\sqrt{m_i^2+r^2\norm{\bm u_i}^2}=M.
 \label{eq:radius-equation}
\end{equation}
Then, the maps from the unit sphere to phase space, and vice versa
\begin{equation}
 \Psi(\bm u)=r_*(\bm u)\bm u,
 \qquad
 \Psi^{-1}(\bm p)=\frac{\bm p}{\norm{\bm p}},
 \label{eq:radial-homeomorphism}
\end{equation}
are mutually inverse homeomorphisms. The resulting topology is a $(3N-4)$-dimensional sphere:
\begin{equation}
 \boxed{\Phi_N(M;\bm m)\cong \hat{S}\cong S^{3N-4}}
 \label{eq:fixed-shell-sphere}
\end{equation}
Intuitively, along each momentum-space ray, the total energy increases monotonically
from $\sum_i m_i$ and reaches $M$ at the unique radius $r_*(\bm u)$. Thus the
spherical direction specifies the complete momentum configuration up to the scale fixed by energy conservation.  We can confirm the dimension of the sphere is correct: we started with $3N$ momentum components, and we have applied $4$ conservation laws. For $N=2$, the shell is perfectly round, and its radius is $r_*=\frac{\sqrt{\lambda}}{\sqrt{2}M}$ where $\lambda$ is the Källén function~\cite{PDGKinematics2025}. For $N\geq3$, the shell is smoothly rounded, albeit ovoidal, when all masses are positive, but can develop sharp features where massless particles have zero momentum, as illustrated schematically in Fig.~\ref{fig:sharp-geometries}.

This homeomorphism transports the round metric, a simple angle on $\hat{S}$ \begin{equation*}
   \theta = \arccos(\bm u\cdot \bm v)
\end{equation*} to phase space
\begin{equation}
 d_{\Phi,\mathrm{rad}}(\bm p,\bm q)
 =\arccos\left(
 \frac{\sum_i\bm p_i\cdot\bm q_i}
 {\norm{\bm p}\norm{\bm q}}\right) 
 \label{eq:fixed-shell-radial-metric}
\end{equation}
The induced
angle is a genuine metric on each fixed-mass shell and
provides a prototype for the radial comparison with variable masses.
The continuity of Eq. \ref{eq:radial-homeomorphism} is physically clear but we formally show it in Appendix~\ref{app:radial-topology}, as well as the metric statement. In principle, the radius could jump between momentum configurations, and we check formally that does not happen.
In principle, there are infinitely many other metrics we could define in this space.

Another particularly natural choice would be the shortest path directly on the shell, in the momentum ambient space, equipped with its Euclidean inner product. Due to the geometry, this would require a geodesic calculation in general. Thus, for event comparison, we use a radial map analogous to Eq.~\eqref{eq:radial-homeomorphism}: normalizing the
combined momentum vector identifies each configuration with a unique point on a unit sphere, so the angle between two such vectors defines a distance on the shell. We next extend the
comparison to events with different reconstructed masses.

\subsection{Adjoining the fixed-mass shells}
\label{sec:collider-shell}

Fixed-mass phase space is appropriate when the identities and masses of the final-state particles are fixed. Reconstructed collider events instead consist of jets or pseudojets whose invariant masses fluctuate from event to event. To compare such events directly, we therefore enlarge phase space by treating each reconstructed mass as a coordinate.

Promoting the masses
from fixed parameters to nonnegative coordinates joins the separate
fixed-mass shells into one fixed-$M$ event space.  Concatenate the $ (\bm p_i,m_i)$ into $\mathsf w=((\bm p_1,m_1),\ldots,(\bm p_N,m_N))$ with $\sum\bm p_i =0$, and define an energy-based norm on this subspace, and the standard Euclidean norm
\begin{equation}
 \norm{\mathsf w}_{E}=\sum^N_{i=1}E_i, \quad  \norm{\mathsf w}=\sqrt{\sum_{j=1}^{4N} \mathsf w^2_j}=\sqrt{\sum^N_{i=1} E_i^2}
 \label{eq:collider-block-norm}
\end{equation}
The signed shell and physical sector are
\begin{align}
 \cC_N^{\mathrm{sgn}}(M)
 &=\{\norm{\mathsf w}_{E}=M\},
 \label{eq:collider-shell-signed}\\
 \cC_N^+(M)
 &=\{\mathsf w\in\cC_N^{\mathrm{sgn}}(M):m_i\geq0\ \text{for every }i\}.
 \label{eq:collider-shell-positive}
\end{align}
Negative $m_i$ are only bookkeeping coordinates for the signed extension. Each fixed-mass phase space is a slice in this enlarged space. 
The available kinetic energy is $M-\sum_i m_i$.  At the threshold
$\sum_i m_i=M$, each fixed-mass shell reduces to the single configuration with all spatial momenta zero.  Masses and momentum
directions therefore cannot be chosen independently throughout this space.

The signed shell has spherical topology as well, directly because $\norm{\cdot}_E$ is a norm.  We are fixing the length of all $\mathsf w$ with respect to this norm, which gives another radial homeomorphism that we don't explicitly need. Its dimension, $(4N-4)$, comes from the dimension of fixed-mass phase space plus $N$ promoted mass variables. 
For $M>0$ and $N\geq2$,
\begin{equation}
 \cC_N^{\mathrm{sgn}}(M)\cong S^{4N-4},
 \qquad
 \cC_N^+(M)\cong B^{4N-4}.
 \label{eq:collider-topologies}
\end{equation}
Here $B^d$ denotes the closed $d$-dimensional ball. Restricting a single coordinate $m_N\geq 0$ cuts the sphere into a closed hemisphere, which is topologically a ball. All remaining $m_i\geq0$ impose further orthant restrictions in the $4N$ dimensional ambient space, but do not change the topology. Appendix~\ref{app:adjoining-topology} gives the complete proof underlying this topology. 

$\cC_N^+(M)$, henceforth the \textit{collider dome}, will be our primary classification space on which to define \textit{angular} distances. Specifically, normalize the complete mass--momentum vector:
\begin{equation}
 \mathsf w\longmapsto\widehat{\mathsf w}
 =\frac{\mathsf w}{\norm{\mathsf w}}
 \label{eq:collider-radial-map-main}
\end{equation}
so $\widehat{\mathsf w}$ lies in the $m_i\geq 0$ section of the unit sphere. At fixed $M$ this map is injective, with image a spherical sector of
dimension $4N-4$. Radial
normalization induces the angular comparison defined in
Section~\ref{sec:collider-metrics}.

\subsection{Convex energy--momentum coordinates}
\label{subsec:convex-collider-ball}
We constructed a space suitable to host momenta with variable nonnegative masses, $\cC_N^+(M)$. However, its ball topology immediately provides us with an additional opportunity: we can perform a \textit{very} familiar non-linear transformation on the whole space to turn it into a convex shape. This convex shape can be equipped with a straight line distance, which will turn out to be a particularly simple choice. These two spaces can then be directly compared in discrimination performance.

Replacing each mass coordinate $m_i$ by the energy $E_i$ makes the total
energy constraint linear. The overall dimension still remains $(4N-4)$, the masses remain non-fixed, but the points are now represented by $(E,p)$ rather than $(p,m)$.  Requiring real nonnegative masses is equivalent to $E_i\geq\norm{\bm p_i}$ for nonnegative energies, so the same events are
described by
\begin{equation}
 \cK_N(M)=
 \left\{(E_i,\bm p_i)_{i=1}^N:
 E_i\geq\norm{\bm p_i},\
 \sum_iE_i=M,\
 \sum_i\bm p_i=0\right\}.
 \label{eq:convex-collider-definition}
\end{equation}
The maps
\begin{align}
 F:\cC_N^+(M)&\longrightarrow\cK_N(M),
 &
 F((\bm p_i,m_i)_i)
 &=\left(\sqrt{\norm{\bm p_i}^2+m_i^2},\bm p_i\right)_i,
 \label{eq:collider-coordinate-homeomorphism}\\
 F^{-1}((E_i,\bm p_i)_i)
 &=\left(\bm p_i,\sqrt{E_i^2-\norm{\bm p_i}^2}\right)_i
 \label{eq:collider-coordinate-inverse}
\end{align}
are continuous inverses.  They are \textit{diffeomorphisms}\footnote{This is a smooth bijection with smooth inverse.} on the all-massive
regions, while the inverse has the expected square-root behavior at a
massless block. Although $F$ preserves the event information, Euclidean distances in
$(E_i,\bm p_i)$ generally differ from angular distances in normalized
$(\bm p_i,m_i)$.  
Unlike the block-norm shell, $\cK_N(M)$ is convex.  If
$A=(E_i,\bm p_i)$ and $B=(E_i',\bm p_i')$ are allowed and
$0\leq t\leq1$,
then conservation is affine and
\begin{equation}
 \norm{(1-t)\bm p_i+t\bm p_i'}
 \leq(1-t)\norm{\bm p_i}+t\norm{\bm p_i'}
 \leq(1-t)E_i+tE_i'.
 \label{eq:collider-convexity}
\end{equation}
Thus the entire chord lies in $\cK_N(M)$.  This region is closed and bounded,
has dimension $4N-4$ after imposing energy--momentum conservation, and
contains $E_i=M/N$, $\bm p_i=0$ in its interior relative to those constraints.
It is therefore a convex realization of the same $B^{4N-4}$ topology. This convex realization is called a \textit{spectrahedron}, which is shown explicitly in
Appendix~\ref{app:gram-geometry}.

For two blocks at $M=1$, conservation leaves
$E_2=1-E_1$ and $\bm p_2=-\bm p_1$, so
\begin{equation}
 \cK_2(1)=\left\{(E_1,\bm p_1):
 \abs{E_1-\frac12}+\norm{\bm p_1}\leq\frac12\right\}.
 \label{eq:collider-bicone}
\end{equation}
This is a four-dimensional solid bicone; restricting the momenta to one
fixed axis gives a filled diamond.  The bicone in the convex panel of
Fig.~\ref{fig:event-geometries} serves as a schematic of the convex
representation for higher-multiplicity events. 

The spectrahedron's straight-line distance is defined in Section~\ref{sec:collider-metrics}.  Straight-line and angular
minimizing paths generally do not map into one another under the change
of representation. 

\subsection{Connection to massless phase-space and spinor frames}
\label{subsec:spinor-connection-main}
We have two spaces suitable for event comparison; this last section does \textbf{not} introduce another one. Here, we make a quick but necessary connection to the spinor literature. 

Setting all masses to zero connects the event space to the spinor and
lightcone descriptions used in the earlier classifier. For $N\geq2$
and $M>0$, the massless event space can be written as
\begin{equation}
 \Phi_N(M;\bm0)\cong V_2(\mathbb C^N)/U(1)^N\cong S^{3N-4}.
 \label{eq:massless-stiefel-sphere}
\end{equation}
Here $V_2(\mathbb C^N)$ consists of pairs of orthonormal complex vectors,
which we call normalized spinor frames. Dividing by $U(1)^N$ removes each
particle's little-group phase (a kinematic redundancy), leaving the physical momentum
shell~\cite{HenningMelia,LM2020}. The lightcone coordinates used in the
earlier classifier~\cite{CCKL2024} describe the region where
$p_i^+=E_i+p_{z,i}>0$ for every particle as
$\mathring\Delta^{N-1}\times S^{2N-3}$. This region contains all events
except those with particles satisfying $E=-p_z$. The product describes only this open region:
at its boundary, the lightcone coordinates break down, and some angular
degrees of freedom become equivalent or collapse.
Appendix~\ref{app:spinor-lightcone} gives the variables and explains how
they relate to the full, closed shell. The same spinor connection extends
to nonzero masses, as developed in Appendix~\ref{app:spinor-frames}. We will only use the product space for comparison with the earlier classifier in Section~\ref{sec:results}.

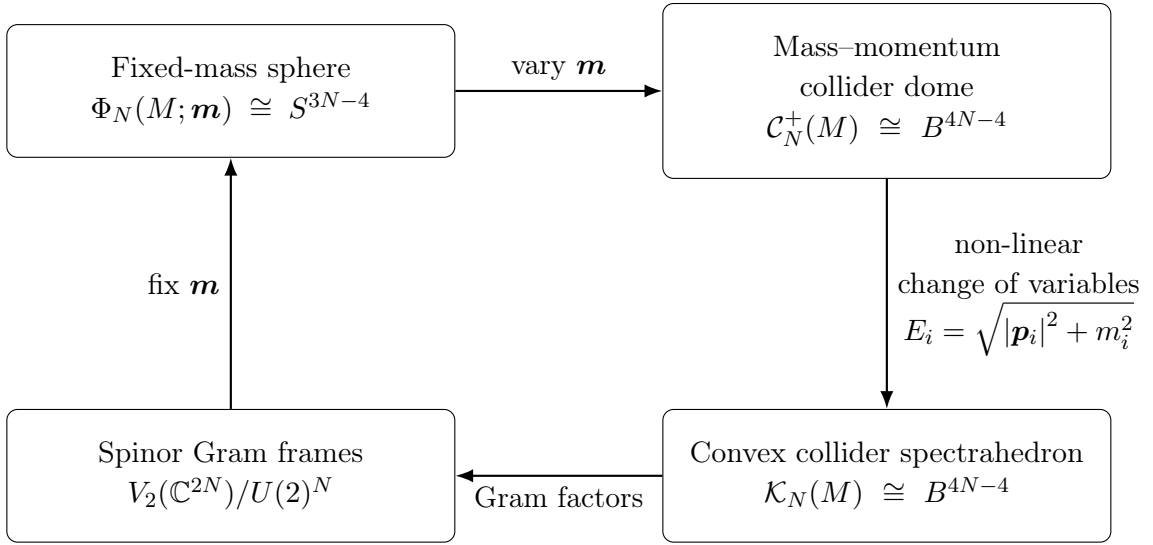
\begin{figure}[H]
\vspace*{0.7\baselineskip}
\centering
\resizebox{\linewidth}{!}{%
\begin{tikzpicture}[
  node distance=3.0cm and 2.7cm,
  box/.style={
    draw,
    rounded corners,
    align=center,
    text width=0.34\linewidth,
    inner xsep=10pt,
    inner ysep=12pt
  },
  arr/.style={-{Latex},thick}
]
\node[box] (fixed) {Fixed-mass sphere\\
  $\Phi_N(M;\bm m)\cong S^{3N-4}$};
\node[box,right=of fixed] (massball) {Mass--momentum collider dome\\
  $\cC_N^+(M)\cong B^{4N-4}$};
\node[box,below=of massball] (convex) {Convex collider spectrahedron\\
  $\cK_N(M)\cong B^{4N-4}$};
\node[box,left=of convex] (stiefel) {Spinor Gram frames\\
  $V_2(\mathbb C^{2N})/U(2)^N$};
\draw[arr] (fixed) -- node[above,align=center]{vary $\bm m$} (massball);
\draw[arr] (massball) -- node[right,align=center]{non-linear\\ change of variables \\$E_i=\sqrt{\norm{\bm p_i}^2+m_i^2}$} (convex);
\draw[arr] (convex) -- node[below,align=center]{Gram factors} (stiefel);
\draw[arr] (stiefel) -- node[left,align=center]{fix $\bm m$} (fixed);
\end{tikzpicture}%
}
\caption{The circle of spaces used in the paper. The arrows record coordinate equivalences, varying the mass parameters, and taking determinant slices; the natural metrics carried by the representations need not agree under these maps. On the massless
determinant slice, the spinor-frame quotient reduces to
$V_2(\mathbb C^N)/U(1)^N$, which is homeomorphic to the massless fixed shell $S^{3N-4}$. The spinor frames do not play a role in classification. }
\label{fig:space-circle}
\end{figure}

Fig.~\ref{fig:space-circle} summarizes this section and the connections between the constructions. 

\section{Representations for event comparison}
\label{sec:collider-metrics}

The mass--momentum and energy--momentum descriptions retain the same events
but assign different distances to them.  We compare four choices: angles
between radially normalized mass--momentum vectors, Euclidean distances
between energy--momentum vectors in the event rest frame, and angles between
two types of energy-weighted features.  Each representation retains the
full event at fixed scale on its stated domain.  Their classification
performance can nevertheless differ because their distances weight
kinematic differences differently.  The metric and pseudometric statements
are proved in Appendix~\ref{app:metricity}.
\subsection{Radial and convex comparisons}

From now on, let $A,B$  denote two collider events. With $\widehat{\mathsf w}=\mathsf w/\norm{\mathsf w}$, radial normalization of the signed shell
gives
\begin{equation}
 d_{\mathrm{coll,rad}}(\mathsf {w}_A,\mathsf w_B)
 =\arccos\ip{\widehat{\mathsf w_A}}{\widehat{\mathsf w_B}}
 =\arccos\left[
 \frac{\sum_i(\bm p_{A,i}\cdot\bm p_{B,i}+m_{A,i} m_{B,i})}
 {\sqrt{\sum_iE_{A,i}^2}
  \sqrt{\sum_iE_{B,i}^2}}
 \right].
 \label{eq:collider-radial-metric}
\end{equation}
At fixed $M$ this is a genuine labeled metric on $\cC_N^{\mathrm{sgn}}(M)$ and the 
physical sector $\cC_N^{+}(M)$.  We call this branch \emph{radial spherical}.

The convex representation carries the particularly simple Euclidean energy--momentum metric
\begin{equation}
 d_{\mathrm E} (A,B)
 =\left[\sum_{i=1}^N
 \left((E_{A,i}-E_{B,i})^2
 +\norm{\bm p_{A,i}-\bm p_{B,i}}^2\right)\right]^{1/2}.
 \label{eq:labelled-fourvector-euclidean}
\end{equation}
It is a genuine labeled metric.
On each fixed-$M$ ball, and likewise on the variable-scale positive cone,
Eq.~\eqref{eq:labelled-fourvector-euclidean} is the intrinsic length distance
induced by the Euclidean line element; its unique constant-speed
minimizing segment is
\begin{equation}
 \Gamma_{A,B}(t)=(1-t)A+tB,\qquad0\leq t\leq1.
 \label{eq:collider-flat-geodesic}
\end{equation}
Thus the convex energy--momentum branch uses the exact straight-line geodesic
metric of the convex collider ball. Related momentum-space event distances, minimized over particle assignments, have been used for
negative-weight reduction in Monte Carlo samples~\cite{AndersenMaier2022}.

The radial and convex constructions thus define distinct metrics on the
same fixed-$M$ physical event space.  The former compares directions in
mass--momentum coordinates, while the latter measures Euclidean separation
in energy--momentum coordinates; both retain the reconstructed masses. Their coordinate relation is given in
Eqs.~\eqref{eq:collider-coordinate-homeomorphism}
and~\eqref{eq:collider-coordinate-inverse}.

\subsection{Moving-frame feature representations}
\label{subsec:moving-frame-features}
The radial spherical distance in
Eq.~\eqref{eq:collider-radial-metric} follows naturally from the
radial homeomorphism. It normalizes the concatenated mass--momentum
vector by a single overall norm, giving greater weight to the
most energetic objects. The topology alone does not require this
weighting. To give subleading kinematic structure greater relative
influence, we construct two alternative spherical representations
that first divide each object's momentum and mass by the square
root of its energy, before applying a common normalization.

Define
\begin{equation}
 \nu_i=\sqrt{\frac{E_i}{M}},\qquad
 x_i=\frac{p_{x,i}}{\sqrt{ME_i}},\quad
 y_i=\frac{p_{y,i}}{\sqrt{ME_i}},\quad
 z_i=\frac{p_{z,i}}{\sqrt{ME_i}},\quad
 \mu_i=\frac{m_i}{\sqrt{ME_i}}. 
 \label{eq:normalized-energy-chart}
\end{equation}
At a zero-energy block, set $\nu_i=x_i=y_i=z_i=\mu_i=0$.
This is continuous because
$x_i^2+y_i^2+z_i^2+\mu_i^2=E_i/M\to0$ as $E_i\to0$. Write $\nu=(\nu_i)$ and similarly $x=(x_i)$, $y=(y_i)$, $z=(z_i)$, and
$\mu=(\mu_i)$.  These variables obey
\begin{gather}
 \sum_i\nu_i^2=1,\qquad
 \sum_i(x_i^2+y_i^2+z_i^2+\mu_i^2)=1,
 \label{eq:energy-chart-spheres}\\
 \nu\cdot x=\nu\cdot y=\nu\cdot z=0.
 \label{eq:energy-chart-orthogonality}
\end{gather}

The three spatial component vectors therefore lie in $\nu^\perp$, which
varies with the event's energy fractions. We place them in a fixed
subspace by choosing a nonnegative unit reference vector $r$ and applying
the Householder reflection
\begin{equation}
 H_r(\nu)=I_N-2\frac{(\nu+r)(\nu+r)^T}{\norm{\nu+r}^2},
 \qquad H_r(\nu)\nu=-r,
 \label{eq:common-householder-frame}
\end{equation}
where $I_N$ is the $N\times N$ identity. The denominator is strictly
positive for nonnegative $\nu$ and $r$. The transformed vectors
$x'=H_r(\nu)x$, $y'=H_r(\nu)y$, and $z'=H_r(\nu)z$ lie in $r^\perp$.
The reflection acts in particle-label space and leaves $\mu$ unchanged;
its orthogonality preserves the unit norm of $(x',y',z',\mu)$.

\paragraph{Democratic representation.}
To treat every object symmetrically, choose
\begin{equation}
 e_{\mathrm{dem}}=\frac{(1,\ldots,1)}{\sqrt N},
 \qquad H_{\mathrm{dem}}(\nu)=H_{e_{\mathrm{dem}}}(\nu).
 \label{eq:democratic-householder}
\end{equation}
Using this reference in the transformed vectors above, define
\begin{align}
 \mathcal H_{\mathrm{dem}}
 &=\left\{(x',y',z',\mu)\in(\R^N)^4:
 e_{\mathrm{dem}}\cdot x'=e_{\mathrm{dem}}\cdot y'
 =e_{\mathrm{dem}}\cdot z'=0\right\},
 \nonumber\\
 \Theta_{\mathrm{dem}}(\mathsf w)
 &=(x',y',z',\mu)
 \in S(\mathcal H_{\mathrm{dem}})\cong S^{4N-4}
 \label{eq:democratic-feature-map}. 
\end{align}
Because all components of $e_{\mathrm{dem}}$ are equal, relabeling the
input objects simply relabels the output features. At fixed $M$, this map
is a homeomorphism from $\cC_N^+(M)$ onto the full spherical sector
$S(\mathcal H_{\mathrm{dem}})\cap\{\mu_i\geq0\}$, including zero-energy
limits. Up to coordinate
ordering, this is the same sector as the radial spherical map's image. The two maps retain
the same event information while generally assigning \textit{different} distances.
 \paragraph{Pivoted representation.}
The pivoted variant tests a comparison in which the highest-$p_T$ object
in each event serves as a distinguished reference, with these leading
objects always matched to one another. For this variant, restrict to
\begin{equation}
 \cC_N^{+,\circ}(M)
 =\{\mathsf w\in\cC_N^+(M):E_i>0\ \text{for every }i\},
 \label{eq:strict-collider-sector}
\end{equation}
and place the unique highest-$p_T$ object in slot $N$. Choose
\begin{equation}
 e_N=(0,\ldots,0,1)^T,
 \qquad H_{\mathrm{piv}}(\nu)=H_{e_N}(\nu).
 \label{eq:pivot-householder}
\end{equation}
The transformed vectors now satisfy $x'_N=y'_N=z'_N=0$. Dropping these
three zero entries gives
\begin{equation}
 \Theta_{\mathrm{piv}}(\mathsf w)
 =(x'_1,\ldots,x'_{N-1};y'_1,\ldots,y'_{N-1};
 z'_1,\ldots,z'_{N-1};\mu_1,\ldots,\mu_N)
 \in S^{4N-4}.
 \label{eq:pivot-feature-map}
\end{equation}
This map is injective on the positive-energy labeled domain with a unique
pivot already in slot $N$.
Highest-$p_T$ ties are excluded from these metric statements, since
choosing a pivot there can introduce ambiguity or discontinuity.

For either moving-frame representation, the angular distance is
\begin{equation}
 d_\Theta(A,B)
 =\arccos\!\left[
 \sum_{i=1}^N
 \left(
 x'_{A,i}x'_{B,i}
 +y'_{A,i}y'_{B,i}
 +z'_{A,i}z'_{B,i}
 +\mu_{A,i}\mu_{B,i}
 \right)\right],
 \qquad
 \Theta\in\{\Theta_{\mathrm{dem}},\Theta_{\mathrm{piv}}\}.
 \label{eq:moving-frame-metric}
\end{equation}
These are metrics on the fixed-$M$ labeled domains specified above (Corollary~\ref{cor:representation-metrics}). On the closed democratic image, the angle is also the shortest-path distance
within the spherical sector; no corresponding claim is made for the
pivoted image. 

Table~\ref{tab:representation-summary} summarizes the representations. The classification pipeline combines these with eventwise
scale normalization, particle assignment, and transverse alignment.  The
resulting pairwise similarities provide the input to the supervised
classifier specified in the following section.
\begin{table}[ht]
\centering
\small
\begin{tabular}{@{}llll@{}}
\toprule
Representation & Compared vector & Distance & Domain \\
\midrule
Radial spherical
  & $\widehat{\mathsf w}$ & Angular & $\cC_N^+(M)$ \\
Convex energy--momentum
  & $(E_i,\bm p_i)_i$ & Euclidean & $\cK_N(M)$ \\
Democratic
  & $\Theta_{\mathrm{dem}}(\mathsf w)$ & Angular & $\cC_N^+(M)$ \\
Pivoted
  & $\Theta_{\mathrm{piv}}(\mathsf w)$ & Angular
  & $\cC_N^{+,\circ}(M)$ \\
\bottomrule
\end{tabular}
\caption{Representations and labeled distances at fixed $M$.}
\label{tab:representation-summary}
\end{table}

\section{Scalable comparison on reconstructed-event space}
\label{sec:implementation}

We have equipped our collider spaces with distances so we are in principle ready for classification. First, we introduce the simulated samples, event selections, and laboratory-frame clustering. We consider three benchmark scenarios: \(t\bar{t}\) versus QCD dijets, four-top production versus \(t\bar{t}+\text{jets}\), and semi-classical black hole signals versus QCD multijets. Next, we adapt the fixed-$M$ constructions of Sections~\ref{sec:fixed-shell} and~\ref{sec:collider-metrics} to events with varying invariant mass by boosting each retained system to its center-of-momentum frame and normalizing its four-momenta to $M=1$. We then describe the sequential assignment--alignment procedure that approximates joint minimization over particle permutations and transverse rotations and reflections, together with its implementation for each representation. Finally, we convert these comparisons into similarities for a support vector classifier and specify the training and evaluation protocol. 
\subsection{Event generation and preprocessing}
\label{subsec:samples}

The input consists of stable final-state particles, with invisible
species excluded by default and no constituent transverse-momentum threshold.
The acceptance cuts in Tables~\ref{tab:samples} and~\ref{tab:blackmax-sample}
define $H_T=\sum_{a\in\mathrm{accepted}}p_{T,a}$.
Accepted particles are clustered in the laboratory frame with exclusive
$k_T$, $R=0.4$, using \textsc{FastJet}~\cite{FastJet}.
Each scan uses one event cohort and the same class balancing at every $N$;
the retained four-vectors are boosted to their center-of-momentum frame and
divided by their invariant mass $M_N$ as in
Eq.~\eqref{eq:eventwise-scale-shape}, with the massless-branch preprocessing
specified in Section~\ref{subsec:representations}.
No zero padding is used: the selected cohorts support the maximum scanned
multiplicity, apart from fewer than $0.1\%$ of $t\bar t$ events with fewer than 30 accepted
particles, which were discarded.  Comparisons at other multiplicities or
with other jet definitions can use the same prescription whenever the
chosen common multiplicity is supported by the sample.

\subsubsection{Standard Model samples}
\begin{sloppypar}
The two Standard Model benchmarks are $t\bar t$ versus QCD dijets and four-top production versus $t\bar t+\text{jets}$, both at $\sqrt{s}=14$~TeV.  The leading-order hard events were generated
with \textsc{MadGraph5\_aMC@NLO} 3.5.13~\cite{MADGRAPH} and showered and hadronized
with \textsc{Pythia}~8.3.16~\cite{PYTHIA}. We work at particle level, providing a
common basis for future detector-response and pileup studies.
Table~\ref{tab:samples} summarizes the samples and selections.
\end{sloppypar}
\begin{table}[!ht]
\centering
\small
\renewcommand{\arraystretch}{1.12}
\begin{tabular}{@{}
 >{\raggedright\arraybackslash}m{0.21\linewidth}
 >{\centering\arraybackslash}m{0.355\linewidth}
 >{\centering\arraybackslash}m{0.355\linewidth}@{}}
\toprule
 & $t\bar t$ versus QCD & $t\bar t t\bar t$ versus $t\bar t+\text{jets}$ \\
\midrule
Collider energy & $\sqrt s=14$ TeV & $\sqrt s=14$ TeV \\
Signal & showered $pp\to t\bar t$ & MadSpin-decayed, all-hadronic $pp\to t\bar t t\bar t$ \\
Background & $pp\to jj$, $j\in\{g,u,d,s,c\}$ and their antiparticles & MadSpin-decayed, all-hadronic MLM-merged $pp\to t\bar t+0,1,2j$, with $j$ including $b$ \\
Generated events & $10000$ signal; $10000$ background & $25000$ signal; $50000$ background \\
Matrix-element selection & dijet $H_{T,j}\geq300$ GeV; inclusive $t\bar t$ & $500\leq H_T^{\mathrm{ME}}\leq1000$ GeV for both classes \\
Particle-level selection & $H_T\geq400$ GeV, $\abs{y}<2$ & $600\leq H_T\leq1000$ GeV, $\abs{y}<2$ \\
Events after selection & $5826$ signal; $6975$ background & $11357$ signal; $19089$ background \\
Balanced classifier sample & $5000$ events per class & $11357$ events per class \\
Multiplicity scan & $N=3,\ldots,15,20,25,30$ & $N=3,\ldots,15,20,25,30,35,40,45$ \\
\bottomrule
\end{tabular}
\caption{Generation and particle-level selections for the two $14$~TeV
benchmarks.  The balanced sample sizes are per-class caps applied before
the stratified training--test split.}
\label{tab:samples}
\end{table}

The top-pair comparison follows the kinematic benchmark of
Ref.~\cite{CCKL2024}, using inclusive showered $pp\to t\bar t$ with
unrestricted top decays.  The dijet generation additionally requires
$p_{T,j}\geq20$~GeV, $\abs{\eta_j}\leq2.5$, $\Delta R_{jj}\geq0.4$,
and leading and subleading parton transverse momenta of at least
$150$ and $100$~GeV. This dataset's classification task provides a relatively simple setting for comparison with the earlier method; unlike in the other two tasks, invisible particles were \textbf{retained} in this sample.
In the four-top comparison, $H_T^{\mathrm{ME}}$ sums transverse momenta of
the non-incoming matrix-element final-state particles before the common
on-shell all-hadronic MadSpin decays~\cite{MadSpin}.
Only the $t\bar t+0,1,2j$ background is $k_T$-MLM matched, with matching
scales $20$~GeV at matrix-element level and $30$~GeV in the shower.
Increasing $N$ resolves additional prongs and radiation and therefore
changes the event representation.

\subsubsection{BlackMax B1 black hole versus QCD multijet sample}
\label{subsec:blackmax-sample}

Extra spatial dimensions allow microscopic black holes to be produced at LHC energies~\cite{ADD1998,DimopoulosLandsberg2001}. At $13$~TeV, we compare the tensionless, nonrotating BlackMax B1
scenario~\cite{Dai_2008} with leading-order QCD multijets.
The signal has Planck mass $M_D=4$~TeV, mass $12\leq M_{\mathrm{BH}}\leq13$~TeV and
$n=6$ extra dimensions; string-ball production, graviton emission, and the modeled mass,
momentum, and angular-momentum losses are disabled.
The background combines $pp\to2j,3j,4j$ with four light quark flavors,
using MLM matching scales of $300$ and $450$~GeV at matrix-element and
shower level, respectively.  Its generation requires
$p_{T,j},m_{jj}\geq300$~GeV, $\abs{\eta_j}<3$, and $H_{T,j}\geq9$~TeV.
Both samples are showered with \textsc{Pythia}~8 using the CP5
tune~\cite{CMS:2019csb}.

The unconditioned samples are strongly separated by global observables.
To reduce this separation, a common $H_T$ window is followed by restrictions
on the $N=30$ invariant mass and linearized transverse sphericity, as
summarized in Table~\ref{tab:blackmax-sample}.

\begin{table}[!ht]
\centering
\small
\renewcommand{\arraystretch}{1.12}
\begin{tabular}{@{}
 >{\raggedright\arraybackslash}m{0.21\linewidth}
 >{\centering\arraybackslash}m{0.355\linewidth}
 >{\centering\arraybackslash}m{0.355\linewidth}@{}}
\toprule
 & BlackMax B1 signal & QCD background \\
\midrule
Collider energy & $\sqrt s=13$ TeV & $\sqrt s=13$ TeV \\
Hard process & BlackMax B1 & $pp\to2j,3j,4j$, $j\in\{g,u,d,s,c\}$ and their antiparticles \\
Common selection & \multicolumn{2}{>{\centering\arraybackslash}m{\dimexpr0.71\linewidth+2\tabcolsep\relax}@{}}{$9.0\leq H_T\leq10.5$ TeV,\quad $\abs{\eta}<2.5$} \\
Particle multiplicity & \multicolumn{2}{>{\centering\arraybackslash}m{\dimexpr0.71\linewidth+2\tabcolsep\relax}@{}}{at least 30 accepted particles} \\
$N=30$ conditioning & \multicolumn{2}{>{\centering\arraybackslash}m{\dimexpr0.71\linewidth+2\tabcolsep\relax}@{}}{$10.5\leq M_{30}\leq11.5$ TeV,\quad $0.10\leq S_{30}\leq0.40$} \\
After all cuts & $6907$ & $8033$ \\
Balanced classifier cap & $6907$ & $6907$ \\
\bottomrule
\end{tabular}
\caption{Selected BlackMax B1 versus QCD sample.  The final row gives
the per-class cap used for classification.}
\label{tab:blackmax-sample}
\end{table}

Here $M_{30}$ is the invariant mass of the sum of the laboratory-frame
$N=30$ objects, and
\begin{equation}
 S_{30}=1-\frac{\sqrt{X^2+Y^2}}{T},\qquad
 T=\sum_i p_{T,i},\quad
 X=\sum_i\frac{p_{x,i}^2-p_{y,i}^2}{p_{T,i}},\quad
 Y=\sum_i\frac{2p_{x,i}p_{y,i}}{p_{T,i}}.
 \label{eq:blackmax-sphericity}
\end{equation}
These cuts define one event cohort, held fixed throughout the multiplicity
scan.  Residual one-variable discrimination is reported in
Appendix~\ref{app:blackmax-numerics}.

\subsection{Eventwise invariant-mass normalization}
\label{subsec:eventwise-normalization}
After particle selection and jet reconstruction, the invariant mass $M$
of the retained system varies from event to event, even at fixed
resolved multiplicity $N$. To compare these events using the
fixed-$M$ constructions above, we boost each retained system to its
COM frame and divide all its four-momenta by its
invariant mass. This gives a common $M=1$ normalization, removing
the overall invariant-mass scale while retaining mass ratios and
relative kinematics.

\begin{figure}[t]
\centering
  \includegraphics[width=0.5\linewidth]
    {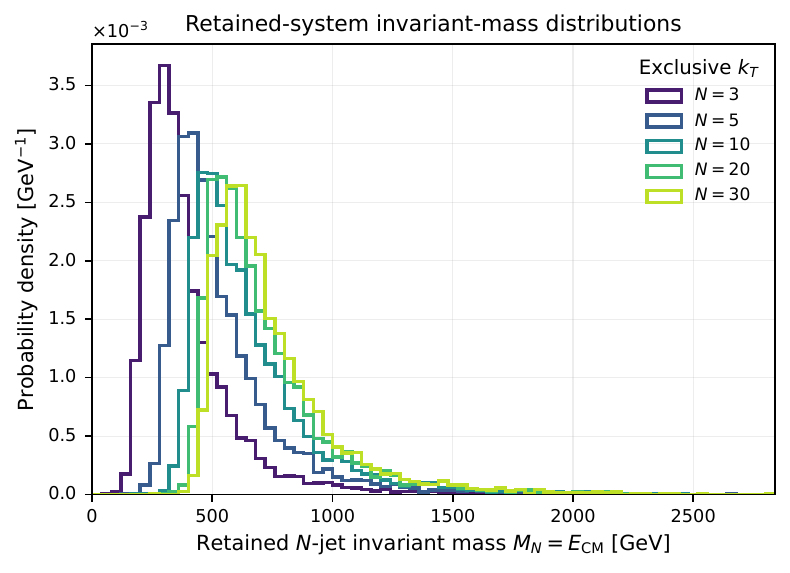}
\caption{Unit-area distributions of the retained-system invariant
mass $M_N$ for the indicated exclusive-$k_T$ multiplicities,
using the first $5{,}000$ accepted events from the generated
$t\bar t$ signal cohort defined in Table~\ref{tab:samples}.
These distributions illustrate the eventwise scale variation
removed by Eq.~\eqref{eq:eventwise-scale-shape}.}
\label{fig:eventwise-mass}
\end{figure}

For each event at exclusive multiplicity $N$, define
\begin{equation}
 P_{N}^\mu=\sum_{i=1}^N p_{N,i}^\mu,
 \qquad
 M_{N}=\sqrt{P_{N}^2}>0.
 \label{eq:event-visible-mass}
\end{equation}
Let $\Lambda_{N}$ boost the retained system to its center-of-momentum
frame, $\Lambda_{N}P_{N}=(M_{N},\bm0)$, and define normalized (\textit{barred}) momenta as
\begin{equation}
 \bar p_{N,i}^\mu
 =\frac{(\Lambda_{N}p_{N,i})^\mu}{M_{N}},
 \qquad
 \sum_{i=1}^N\bar p_{N,i}^\mu=(1,\bm0).
 \label{eq:eventwise-scale-shape}
\end{equation}
Fig.~\ref{fig:eventwise-mass} shows the distribution of $M_N$
for a representative $t\bar t$ signal cohort. The
retained-system invariant mass is broadly
distributed at every displayed $N$, with a median that increases from
$340$~GeV at $N=3$ to $675$~GeV at $N=30$. Because the
hadron-collider exclusive-$k_T$ history includes beam assignments, the
four-vector sum of the retained jets, and hence $M_{N}$, also need not be independent of $N$.  The division in
Eq.~\eqref{eq:eventwise-scale-shape}, as well as the center-of-momentum
boost, selects a common unit-$M$ representative.

All numerical comparisons below use the $M=1$ representative.
This normalization
sends $\mathsf w_A$ to $\mathsf w_A/M_A\in\cC_N^+(1)$,
or equivalently places their energy--momentum coordinates in
$\cK_N(1)$. A fixed-mass shell $\Phi_N(M;\bm m)$ is sent to
$\Phi_N(1;\bm m/M)$. Because these mass ratios vary across events,
the normalized sample lies in the variable-mass collider spaces
rather than one fixed-mass sphere.

While the fixed-scale labeled distances remain metrics on the normalized image, their pullbacks to the COM event
family are \textit{pseudometrics}, identifying events that differ only by a common positive rescaling. This means that events that only vary by an overall scale are distance zero. One can imagine that this affects events that are \textit{close to an overall scale apart as well}: events with similar normalized kinematics can lie close together even when their invariant-mass scales differ substantially. Thus, the loss of scale information affects event proximity beyond \textit{exactly} rescaled copies. For example, let \(A\) contain four equal-energy massless particles arranged in two orthogonal back-to-back pairs, and obtain \(B\) by rescaling the pairs by \(\lambda(1+\varepsilon)\) and \(\lambda(1-\varepsilon)\), respectively. Although \(M_B=\lambda M_A\), their normalized distances are
\begin{equation}
    d_{\mathrm E}(\bar A,\bar B)=\frac{|\varepsilon|}{\sqrt2},
\qquad
d_{\mathrm{coll,rad}}(\bar A,\bar B)=\arctan|\varepsilon|.
\end{equation}
Thus, approximately rescaled events can have arbitrarily small distances regardless of their overall scale separation. The probability of this for a given pair goes roughly like\begin{equation}
    P(\norm{A-B}<\varepsilon) \sim \varepsilon^{4N-4}
\end{equation}
and over all the pairs across $n_{\text{events}}$,
\begin{equation}
    P(\text{At least 1 pair is within $\varepsilon$}) \sim \frac{n_\text{events}^2}{2}\varepsilon^{4N-4}
\end{equation}
and so we leave studying this particular issue deeply to further work.

\subsection{Exact assignment target versus implemented dissimilarity}

\label{subsec:exact-vs-production}
Events described only by reconstructed four-momenta have no
preferred particle labeling, while the beam axis singles out
the longitudinal direction but no preferred azimuthal orientation.
Following Ref.~\cite{CCKL2024}, we compare events up to allowed
particle permutations and transverse rotations and reflections,
represented by $O(2)$. Exhaustively minimizing over all $N!$
assignments and their transverse alignments becomes impractical
at high multiplicity, motivating the scalable sequential
approximation described below.

For normalized events $A,B$, let $d_{\mathrm{rep}}$ denote the
chosen labeled distance and $G$ the allowed permutation group:
$S_N$ for unrestricted matching or $S_{N-1}$ when a distinguished
pivot is held fixed. The exact distance between symmetry orbits $[A]$ and $[B]$ is
\begin{equation}
 D_{\mathrm{joint}}([A],[B])
 =\min_{\gamma\in G\times O(2)}
 d_{\mathrm{rep}}(A,\gamma\cdot B).
 \label{eq:event-joint-target}
\end{equation}
This is a metric on the corresponding normalized
representation-orbit space, as established in
Appendix~\ref{app:exact-metrics}. This applies to all Section~\ref{sec:collider-metrics} representations.

We approximate this joint minimum in two steps. First, the
Hungarian algorithm~\cite{Kuhn1955} selects an assignment $\sigma_H$ in the
input frame,
\begin{equation}
 \sigma_H\in\arg\min_{\sigma\in G}\sum_i C_{i,\sigma(i)},
 \label{eq:fixed-frame-assignment}
\end{equation}
where $C_{ij}$ is the cost of matching block $i$ in $A$ to block
$j$ in $B$. The representation-specific costs are given in
Section~\ref{subsec:representations}. 

We then minimize over transverse alignment with this assignment
held fixed:
\begin{equation}
 D_{H\to O(2)}(A,B)
 =\min_{R\in O(2)}
 d_{\mathrm{rep}}(A,R\cdot(\sigma_H B)).
 \label{eq:two-stage-dissimilarity}
\end{equation}
The Hungarian step is exact for the chosen fixed-frame additive
objective, and the $O(2)$ step is exact for the selected assignment.
However, the fixed-frame optimal assignment need not minimize
the distance after alignment. Their composition therefore gives
a polynomial-time dissimilarity\footnote{We will keep using the term distance for ease of reading, despite the lack of metric guarantee.} on the input representatives,
without the metric guarantee of the exact joint minimum.

For a fixed assignment in a spherical representation, collect transverse
coordinates into $\xi_i=x_i+iy_i$ and retain real longitudinal and mass-like
coordinates $z_i,\mu_i$.\footnote{For transformed branches, this step is done on $x',y',z'$.}   Define
\begin{equation}
 Z_\sigma=\xi_A^\dagger \xi_{B,\sigma},
 \qquad T_\sigma=\xi_A^T \xi_{B,\sigma}.
\end{equation}
The two quantities encode the rotational and reflection branches of $O(2)$.
The exact fixed-assignment cosine is
\begin{equation}
 \cos\theta_\sigma=
 \frac{\max(\abs{Z_\sigma},\abs{T_\sigma})
 +z_A\cdot z_{B,\sigma}+\mu_A\cdot \mu_{B,\sigma}}
 {\sqrt{\xi_A^\dagger \xi_A+\norm{z_A}^2+\norm{\mu_A}^2}
  \sqrt{\xi_B^\dagger \xi_B+\norm{z_B}^2+\norm{\mu_B}^2}}
 \label{eq:fixed-permutation-cosine}
\end{equation}

Appendix~\ref{app:assignment} quantifies the approximation for all four mass-aware
branches at $N\leq 10$, comparing the sequential procedure
with exhaustive minimization over each branch's allowed
permutations and exact transverse alignment. 
\subsubsection{Compared representations}
\label{subsec:representations}

We compare five representations built from the same reconstructed events. Four retain reconstructed masses: the convex
energy--momentum representation uses a Euclidean distance, while the radial and two moving-frame representations use angular distances. The massless lightcone representation provides a
benchmark against the earlier method~\cite{CCKL2024}.
For each representation, we specify the cost matrix used in the assignment--alignment procedure described above. The massless
benchmark is also evaluated using its original rank ordering.

For either moving-frame branch, write
$b_i'=(x_i',y_i',z_i',\mu_i)$.

\begin{description}[leftmargin=2.6em,style=nextline]
\item[Radial spherical.]
Use the normalized physical blocks
$\bar b_i=(\bar p_{x,i},\bar p_{y,i},\bar p_{z,i},\bar m_i)$, with
$\bar m_i=m_i/M\geq0$ on physical data, in
Eq.~\eqref{eq:collider-radial-metric}.  The fixed-frame assignment cost is
$C_{ij}^{\mathrm{rad}}=-\bar b_{A,i}\cdot\bar b_{B,j}$.  Hungarian assignment first
computes the additive permutation target exactly; the subsequent alignment
gives the implemented two-stage distance in
Eq.~\eqref{eq:two-stage-dissimilarity}.

\item[Pivoted moving frame.]
Use $\Theta_{\mathrm{piv}}$ from Eq.~\eqref{eq:pivot-feature-map}, hold the
highest-$p_T$ block fixed, and assign the remaining $N-1$ blocks using
$C_{ij}^{\mathrm{piv}}=-b'_{A,i}\cdot b'_{B,j}$.  This defines the pivoted
branch.

\item[Democratic moving frame.]
Use the permutation-equivariant map defined by
Eq.~\eqref{eq:democratic-feature-map}, with Householder vector
$\nu+e_{\mathrm{dem}}$, and assign all $N$ transformed blocks with
$C_{ij}^{\mathrm{dem}}=-b'_{A,i}\cdot b'_{B,j}$.  This same convention is
used in all three production studies.

\item[Convex energy--momentum.]
For normalized events $\bar A,\bar B$, a fixed assignment, and $R\in O(2)$, define
$\widetilde R=\operatorname{diag}(R,1)\in O(3)$.  Then
\begin{equation}
 d_{\mathrm E}^2(\bar A,\bar B;\sigma,R)
 =\sum_i(\bar E_{A,i}-\bar E_{B,\sigma(i)})^2
 +\sum_i\norm{\bar{\bm p}_{A,i}
 -\widetilde R\bar{\bm p}_{B,\sigma(i)}}^2.
 \label{eq:normalized-convex-distance}
\end{equation}
The fixed-frame additive cost
\begin{equation}
 C_{ij}^{\mathrm E}
 =(\bar E_{A,i}-\bar E_{B,j})^2
 +\norm{\bar{\bm p}_{A,i}-\bar{\bm p}_{B,j}}^2
\end{equation}
is minimized exactly by Hungarian assignment.
For this branch set
$\xi_i^{\mathrm E}=\bar p_{x,i}+i\bar p_{y,i}$,
$Z_\sigma^{\mathrm E}=(\xi_A^{\mathrm E})^\dagger
\xi_{B,\sigma}^{\mathrm E}$, and
$T_\sigma^{\mathrm E}=(\xi_A^{\mathrm E})^T\xi_{B,\sigma}^{\mathrm E}$.
At fixed $\sigma$, minimizing the transverse term is equivalent to maximizing
the transverse cross term
$\max(\abs{Z_\sigma^{\mathrm E}},\abs{T_\sigma^{\mathrm E}})$.  The labeled
distance is the straight-chord distance of
Eq.~\eqref{eq:labelled-fourvector-euclidean}; the implemented
assignment--alignment composition remains sequential.
\item[Massless lightcone benchmark (Rank-ordered and Hungarian).]
For both massless comparisons, we follow the event representation and
preprocessing of Ref.~\cite{CCKL2024}, using the original implementation. After exclusive-$k_T$ clustering, each block is
projected to the massless shell by replacing
\(
E_i\mapsto\norm {\bm p_i}
\).
The resulting massless event is then boosted to its own
COM frame and normalized by its total COM
energy before the simplex--sphere coordinates of
Refs.~\cite{LM2020,CCKL2024} are constructed.

The legacy baseline orders the blocks by decreasing $p_T$.
Both massless branches use the plus lightcone convention,
$p_i^+=E_i+p_{z,i}$, for the simplex coordinates. The
legacy curves use sphere-to-simplex weights $c=0.25,1,4$.
The full implementation and coordinate definitions for
$(\rho,v)\in\mathring{\Delta}^{N-1}\times S^{2N-3}$
are given in Ref.~\cite{CCKL2024}.

The massless lightcone--Hungarian branch leaves the massless projection,
boost, normalization, lightcone convention, weighting,
and final transverse alignment unchanged. It replaces rank ordering
with fixed-frame Hungarian assignment and uses the highest-$p_T$ block
as its pivot. The legacy baseline instead uses the lowest-$p_T$ block,
which occupies the final slot under descending ordering.
On the non-pivot labels, the assignment cost is
\begin{equation}
 C_{ij}^{\mathrm{ml}}
 =\alpha(\rho_{A,i}-\rho_{B,j})^2
 +2\beta\left[
 1-\operatorname{Re}\!\left(v_{A,i}^*v_{B,j}\right)
 \right],
 \qquad
 \alpha,\beta>0,
 \qquad
 \frac{\beta}{\alpha}=c.
 \label{eq:massless-hungarian-cost}
\end{equation}
After selecting $\sigma_H$ with this cost, we evaluate the same
fixed-assignment simplex--sphere distance as in the legacy implementation,
including exact minimization over the common transverse $O(2)$ action, as
in Eq.~\eqref{eq:two-stage-dissimilarity}. 
The massless lightcone-Hungarian branch is included primarily to benchmark the new representations against the approach of Ref.~\cite{CCKL2024}. To limit the computational cost of this comparison, we fixed $c$ separately for each task to the value yielding the best performance in pilot runs, without repeating the full analysis for all three values, $c=0.25,1,4$.
\end{description} 

\subsection{Classifier and precomputed similarities}
\label{subsec:classifier}

For each distance matrix $D$, the implementation forms
\begin{equation}
 K_{ab}=\exp(-\gamma D_{ab}^2)\quad\text{(all operations element-wise)}
 \label{eq:rbf-similarity}
\end{equation}
with $\gamma>0$, and supplies it as a precomputed kernel to a support vector
classifier~\cite{CortesVapnik,ChangLin2011}. The metric properties of the idealized event comparisons do not
by themselves guarantee positive semidefiniteness of these Gaussian
similarities. A sufficient condition, and a necessary one for such
a guarantee at every $\gamma>0$, is conditional negative
semidefiniteness\footnote{CND is 
$\sum_{i,j}c_i c_j d(x_i,x_j)^2\leq 0$
for every finite choice of points and real coefficients
satisfying $\sum_i c_i=0$.} of $D^2$.
Squared geodesic distances on curved manifolds do not generally
satisfy this condition~\cite{Feragen2015}.

We have not systematically verified positive semidefiniteness of
the training kernels produced by the implemented
assignment--alignment comparisons; for some diagnostic cases, particularly at low $N$, negative eigenvalues were observed. A particularly extreme instance of this was for an $N=5$ democratic sphere kernel where $\frac{|\lambda_{\text{min}}|}{\lambda_{\text{max}}}\approx5\%$. Because the Hessian of the SVC dual minimization objective has the same eigenvalues as the kernel matrix, an indefinite kernel can make the optimization nonconvex, so the solver's stopping criterion need not certify a global optimum~\cite{ChangLin2011}. Nevertheless, classification performance was evaluated by applying the trained classifiers to unseen, held-out events. The reported test AUCs (\textbf{A}rea \textbf{U}nder receiver operating characteristic \textbf{C}urve) therefore measure the discrimination actually achieved by the fitted models, including any effects of convergence to a suboptimal solution. The absence of a convexity guarantee leaves open whether a better value of the training objective could have been obtained, but it does not invalidate the evaluation of the resulting classifier on independent test data. Global optimality of the training solution is not required for this assessment of predictive performance.

The data are split with stratification into $70\%$ training and $30\%$
test sets.  For each outer training split, set
\begin{equation}
 \gamma_0=
 \frac{\log 2}
 {\operatorname{median}_{a<b,\;a,b\in\mathrm{train}}D_{ab}^2},
 \qquad
 C_{\mathrm{SVC}}\in\{0.03,0.1,0.3,1,3,10\}.
\label{eq:gamma-grid}
\end{equation}
The common bandwidth grid is
\begin{equation}
 \gamma/\gamma_0\in\{0.3,1,3,10,20,30\}.
 \label{eq:numerical-gamma-grid-main}
\end{equation}
The massless lightcone--Hungarian weight is $c=1$ for $t\bar t$ and
BlackMax and $c=0.25$ for four-top; the massless grids have task and branch specific extensions.\footnote{In the $t\bar t$ assignment comparison, one common outer
split for the legacy and Hungarian lightcone branches also includes
$\gamma/\gamma_0=0.03,0.1$; the remaining splits use the common grid. These extensions were motivated by pilot calibration runs that frequently
selected the lower boundary of the initial grid.}
For each outer split, five independently shuffled three-fold
cross-validations, with inner seeds $0$--$4$, each select a
$(C_{\mathrm{SVC}},\gamma)$ pair by maximizing the mean validation AUC over
its three folds; each selected pair is refitted on the full outer training
sample and evaluated on the same held-out test sample.
For a given outer seed, the reported AUC is the held-out test AUC of the
candidate with the highest mean validation AUC among these five inner runs. In the extremely rare case of validation AUC ties, the two or more test AUCs are computed and averaged.
Each plotted point is the mean of these selected test AUCs over outer seeds
that generate new stratified training--test splits, including a new held-out
sample. Its error bar is the standard error across outer splits, to quantify variation across repeated outer splits.

Once trained, the SVM classifies a new event by comparing it with a subset of the training events called support vectors. The distance to each support vector is converted into a kernel similarity using the selected $\gamma$, and these similarities are combined with learned coefficients and a bias to obtain a decision score. With signal assigned the positive label, the sign determines the predicted class, while larger scores indicate greater signal likeness. Under the positive semidefinite kernel assumption, this score is proportional to the signed distance from the learned decision hyperplane in the associated feature space.

\section{Classification results}
\label{sec:results}
The same reconstructed event is compared here through several coordinate and
feature representations of the unit-$M$ collider spaces.  The convex
energy--momentum, radial, and energy-adapted moving-frame constructions
retain the complete normalized massive event on their stated domains, with
the moving frames taking values in spheres; the massless lightcone branch connects to the
earlier pipeline.  The $t\bar t$ section first compares the legacy and updated
massless prescriptions, after which all five representations are compared
on the two Standard Model samples and the conditioned BlackMax sample.
Since the coordinate or feature map, treatment of mass coordinates,
and assignment cost vary together, the AUC curves rank the complete
implemented representations.  The democratic moving frame has the highest
mean AUC at the endpoint of all three scans, with process-dependent
crossovers and saturation as the resolution is increased.
\subsection{\texorpdfstring{$t\bar t$}{ttbar} versus QCD dijets}
\label{sec:ttbar-results}

\begin{figure}[!ht]
    \centering
    \includegraphics[width=0.88\linewidth]
      {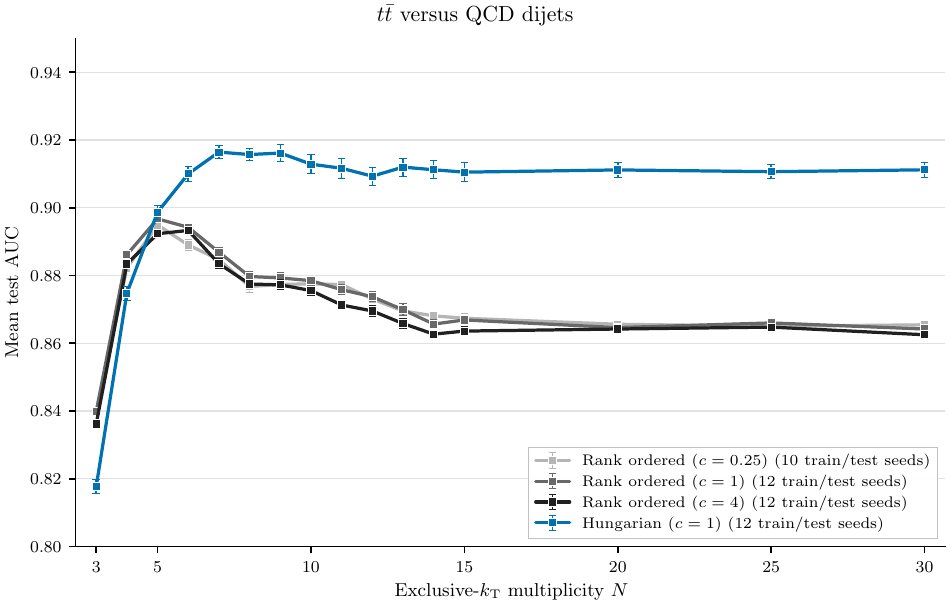}
    \caption{Mean test AUC versus exclusive-$k_T$ multiplicity for the
    showered \(pp\to t\bar t\) sample and QCD dijets, comparing the
   ordering prescription of Ref.~\cite{CCKL2024} at \(c=0.25,1,4\)
    with fixed-frame Hungarian assignment in the same lightcone parametrization.
    Points are means over the training--test splits indicated in the legend,
    and bars are standard errors.}
    \label{fig:ttbar-assignment}
\end{figure}

The $t\bar t$ benchmark is a natural choice to compare the present construction to the
event-level classification of Ref.~\cite{CCKL2024}.
Figure~\ref{fig:ttbar-assignment} compares the legacy lightcone ordering
with the updated massless prescription, which combines Hungarian
assignment with a highest-$p_T$ pivot and subsequent transverse alignment.
The main change is the behavior at high multiplicity. The legacy
curves reach their maxima around $N=5$--$6$ and then lose discrimination,
whereas the updated branch reaches a broad maximum around $N=7$--$9$ and
retains most of that performance through $N=30$. At matched weight $c=1$,
the endpoint gain is $4.70\pm0.20$ AUC percentage points over the 12 common
outer splits. Thus, the combined update to assignment and pivot choice
substantially improves the use of finely resolved events within the
massless construction.

\begin{figure}[!ht]
    \centering
    \includegraphics[width=0.88\linewidth]
      {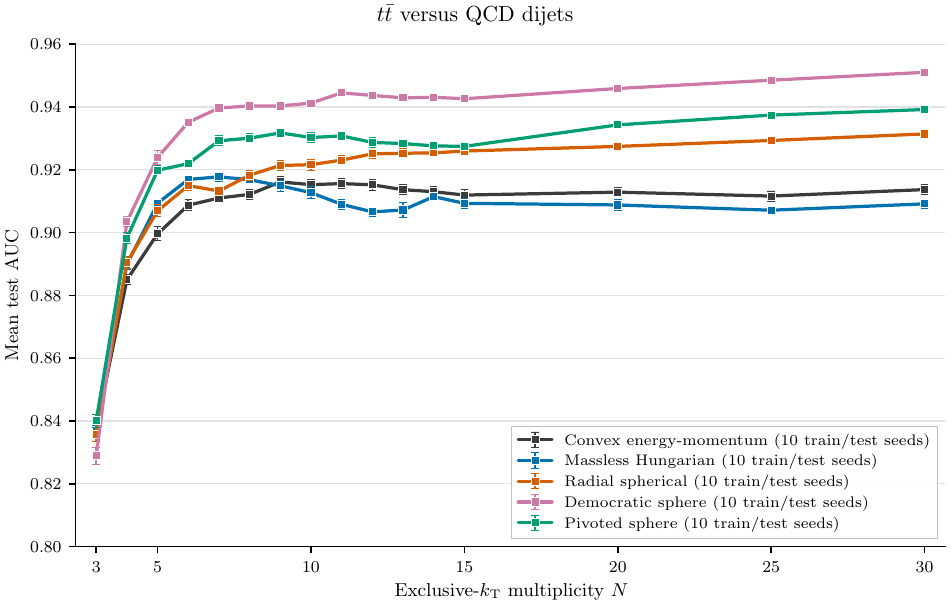}
    \caption{Mean test AUC versus exclusive-$k_T$ multiplicity for the
    showered \(pp\to t\bar t\) sample and QCD dijets, comparing the five
    branches defined in Sec.~\ref{subsec:representations}.
    All branches use ten outer splits.
    Points are outer-split means and bars are standard errors.}
    \label{fig:ttbar-qcd}
\end{figure}

The comparison of all five pipelines in Fig.~\ref{fig:ttbar-qcd} shows a
further gain from the moving-frame representations. All branches improve
rapidly as the event is first resolved into more blocks. The pivoted frame
has the highest mean at $N=3$, and the democratic frame leads at every
scanned multiplicity from $N=4$ onward. Much of its discrimination is
already present at modest resolution: its mean AUC is
$0.9397\pm0.0014$ at $N=7$. After a broad intermediate-resolution region,
it improves further to $0.9510\pm0.0009$ at $N=30$.
This comparison uses ten common outer splits across all five branches.\footnote{
The massless lightcone--Hungarian curve in Fig.~\ref{fig:ttbar-qcd}
uses a different set of outer splits from the assignment comparison in
Fig.~\ref{fig:ttbar-assignment}.}

The high-resolution ordering distinguishes all five representations.
The pivoted frame remains second throughout the scan from $N=4$ onward,
and the radial spherical branch also benefits from finer resolution.
From $N=8$ onward, all three massive spherical branches exceed both the
convex and massless branches in mean AUC. The convex energy--momentum
comparison saturates near its intermediate-resolution maximum, while
the massless branch settles below its earlier peak. At $N=30$, the
democratic advantage over the pivoted frame is $1.19\pm0.06$ AUC
percentage points, and the gaps to the convex and massless branches
are approximately four points. The progression from the assignment
comparison to the complete-pipeline comparison therefore identifies two
useful improvements: greater stability with increasing multiplicity in
the updated massless prescription, and stronger discrimination from the
mass-aware spherical representations at finer resolution.
\subsection{Fully hadronic four-top versus merged \texorpdfstring{$t\bar t+\text{jets}$}{ttbar+jets}}
\label{subsec:fourtop-results}

We next consider a substantially more challenging
event-level benchmark: fully hadronic four-top production against a merged
top-pair-plus-jets background.  The selected samples and common
particle-level processing are summarized in Table~\ref{tab:samples}.

The \(t\bar t+\text{jets}\) sample is a challenging background with
substantial signal-like structure: it already contains real tops, heavy flavor, hadronic
\(W\) decays, and hard radiation.  The benchmark therefore measures
full-pipeline discrimination in this more closely matched process-level setting; dedicated
scalar baselines would quantify the remaining coarse-observable contribution
separately.  In the merged background, up to two additional hard emissions
are described at matrix-element level, with MLM vetoes preventing double
counting with the shower.  Since fully hadronic four-top production has an
intrinsic twelve-parton core before showering, it motivates testing finer
exclusive-$k_T$ resolutions over the wider range \(3\leq N\leq45\).

\begin{figure}[!ht]
    \centering
    \includegraphics[width=\linewidth]
      {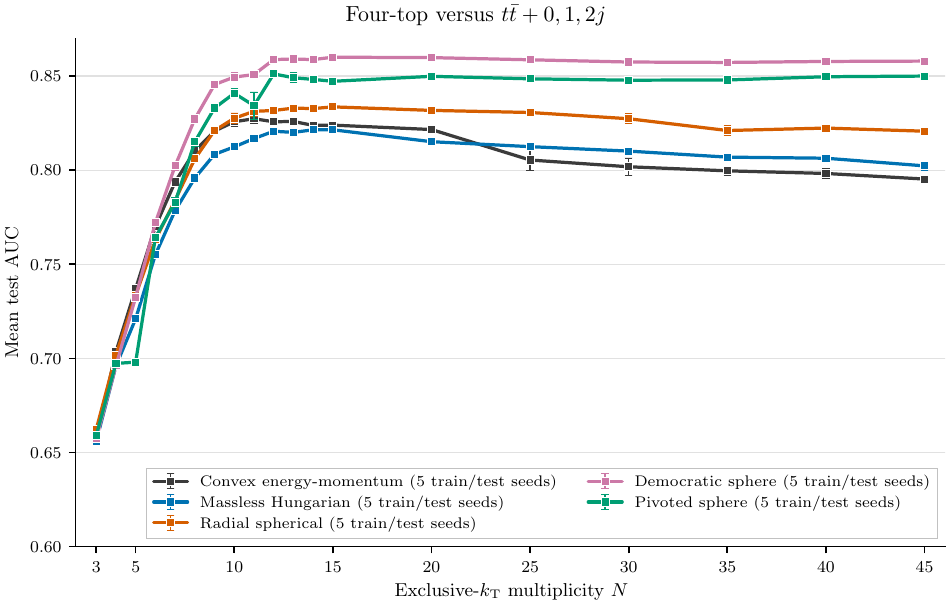}
    \caption{Mean test AUC versus exclusive-\(k_T\) multiplicity for fully
    hadronic \(t\bar t t\bar t\) versus MLM-merged
    \(t\bar t+0,1,2j\).  The curves compare the convex energy--momentum,
    massless lightcone--Hungarian, radial spherical, democratic moving-frame,
    and pivoted moving-frame representations.  Points are means over five
    repeated \(70{:}30\) outer training--test splits shared by all branches, and
    bars are their standard errors.}
    \label{fig:fourtop}
\end{figure}

The main feature of Fig.~\ref{fig:fourtop} is the separation between the
two moving frames and the other three branches as the resolution
increases. At the coarsest resolutions, the radial branch leads at $N=3$
and the convex branch at $N=4$--$5$. The democratic frame takes the lead
at $N=6$, followed by the pivoted frame in second place from $N=8$ onward.
Both moving frames reach broad plateaus around $N=12$--$15$ and sustain
their discrimination through $N=45$. The democratic branch attains its
largest scanned mean, $0.8600\pm0.0018$, at $N=15$ and ends at
$0.8579\pm0.0013$. Its advantage is therefore already present at
intermediate resolution and persists across the remainder of the scan. The notable \(N=11\) dip and error bar in the pivot branch are due to two of the five outer splits yielding markedly lower AUCs than the others. This behavior is consistent with increased sensitivity to the choice of pivot near the twelve-parton decay multiplicity, where several resolved hard prongs may have comparable transverse momenta.

The radial, convex, and massless branches instead lose performance beyond
their intermediate-resolution maxima. The radial branch remains the
strongest of these three at high multiplicity. The convex comparison,
despite its early lead, shows the largest subsequent decline and falls
below the massless branch from $N=25$ onward. By $N=45$, the democratic
frame exceeds the convex and massless branches by $6.27\pm0.13$ and
$5.57\pm0.15$ AUC percentage points, respectively, while its advantage
over the pivoted frame is $0.80\pm0.04$ points. This benchmark makes the
resolution dependence particularly clear: the moving frames preserve
their classification performance as the event is resolved at multiplicities well above
the hard decay multiplicity, whereas the other representations favor a
coarser comparison.

This benchmark also reaches the highest multiplicities considered here. At these resolutions, reduced fidelity of the sequential assignment--alignment approximation could contribute to the decline observed in some branches. Comparing both distance fidelity and classification performance with more accurate joint minimization would help determine the contribution of the approximation to this resolution dependence.

The comparison with the radial branch also connects directly to the
motivation in Section~\ref{subsec:moving-frame-features}. Dividing each object's momentum
and mass by the square root of its energy gives subleading kinematic
structure greater relative influence than radial normalization. Their sustained performance
at high multiplicity is consistent with that motivation. The radial and
democratic maps preserve the same normalized kinematic information and
the same spherical-sector topology, so their different AUC curves
illustrate the importance of how this information enters the event
comparison. Controlled variations of the weighting, mass treatment, and
allowed assignments would quantify their individual contributions.

\subsection{Conditioned BlackMax B1 versus QCD multijets}
\label{subsec:blackmax-results}

Massless phase-space-distance SVM classification has already been applied to BlackMax
signals in the CMS black-hole and sphaleron analysis~\cite{CMSBlackHoles2026}. The conditioned BlackMax B1 sample defined in Section~\ref{subsec:blackmax-sample} provides a BSM-versus-four-flavor-QCD stress test at substantially higher visible energy than the Standard Model benchmarks. The \(N=30\)
conditioning defines one event cohort, which is then held fixed and
reclustered across \(N\).  Before the
\(M_{30},S_{30}\) conditioning, substantial scalar separation remains even
within the common \(H_T\) window.  The additional restrictions reduce it to
the residual structure quantified here.  On the conditioned sample before
balancing, the largest orientation-adjusted one-variable AUC is \(0.7701\)
for the
laboratory-frame block-energy sum
\(E_{\mathrm{vis}}^{\mathrm{lab}}=\sum_{i=1}^{30}E_i^{\mathrm{lab}}\), followed by
\(0.7586\) for \(S_{30}\) and \(0.7471\) for \(M_{30}\); the summed
mass of the \(N=30\) blocks gives \(0.5279\).  These direct one-variable scores provide
scalar reference points for this comparison; a matched multivariate scalar
model is a natural next comparison.

\begin{figure}[!ht]
 \centering
 \includegraphics[width=\linewidth]
   {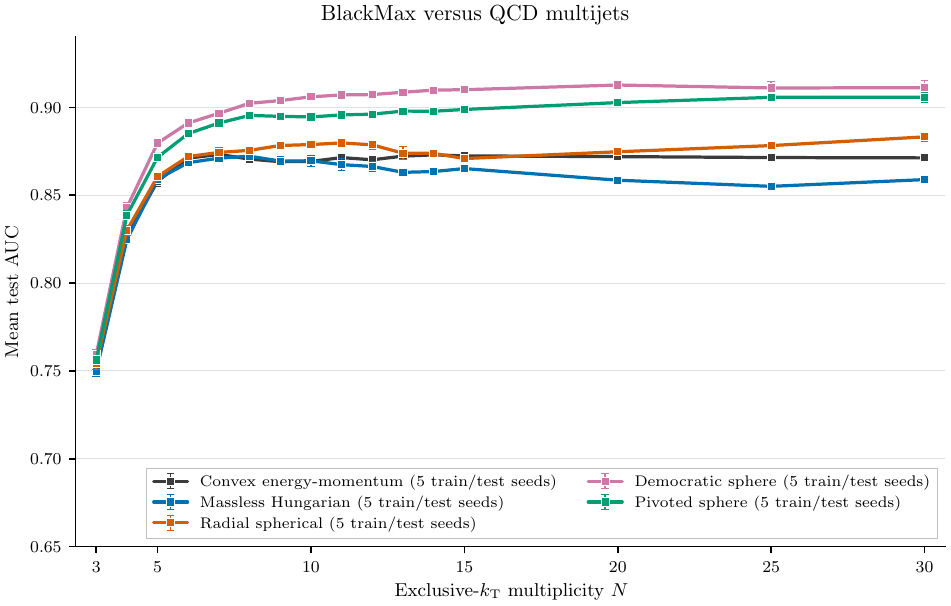}
 \caption{Mean test AUC versus exclusive-\(k_T\) multiplicity for BlackMax
 B1 versus four-flavor QCD multijets.  The curves compare the convex
 energy--momentum, massless lightcone--Hungarian, radial spherical, democratic
 moving-frame, and pivoted moving-frame representations.  Points are means
 over the five common \(70{:}30\) outer splits, and
 bars are their standard errors.}
 \label{fig:blackmax-qcd4f}
\end{figure}

The moving frames lead throughout this scan, with the democratic branch
first and the pivoted branch second at every multiplicity in
Fig.~\ref{fig:blackmax-qcd4f}. Both continue to improve beyond the
resolutions at which the convex and massless branches reach their
largest means. The democratic curve forms a broad high-resolution
plateau, with a maximum of $0.9127\pm0.0016$ at $N=20$ and an endpoint
of $0.9112\pm0.0039$ at $N=30$. The pivoted frame draws closer at the largest
multiplicities; their paired endpoint difference is
$0.55\pm0.29$ AUC percentage points. The two moving frames consequently
provide comparable high-resolution performance on this sample, with a
consistent lead in mean AUC for the democratic construction.

The remaining branches show three distinct responses to finer
resolution. The convex comparison settles into an early plateau,
the massless comparison declines after its maximum, and the radial
comparison gains again at high multiplicity to finish between the
moving frames and the other two branches. At $N=30$, the democratic
frame exceeds the convex and massless branches by $3.99\pm0.23$ and
$5.23\pm0.39$ AUC percentage points, respectively. Together with the
Standard Model benchmarks, this shows a recurring advantage of
the moving-frame pipelines across the selected samples, accompanied by
process-dependent gains from finer resolution: continued improvement
for $t\bar t$, and broad plateaus for four-top and conditioned BlackMax.

\section{Conclusions and future outlook}
\label{sec:conclusions}

We have extended phase-space-based event comparison to variable
reconstructed masses and scalable particle assignment for supervised
binary classification.  The known spherical topology of fixed-mass phase
space supplies the starting point.  An explicit radial map defines angular
distances, and treating masses as coordinates allows events from different
fixed-mass shells to be compared.  The radial, convex, pivoted, and democratic
representations retain the normalized four-vectors on their stated domains
while assigning different event distances.  In particular, the radial
and democratic constructions give distinct homeomorphisms of the closed
labeled collider space onto the same spherical sector, preserving its
kinematic information and topology.  The appendices relate these
representations to spinor frames and the physical phase-space measure.

Eventwise normalization removes the common invariant-mass scale.
The fixed-scale labeled metrics remain metrics on the normalized spaces;
when applied to variable-scale events, these are pseudometrics.
Exact simultaneous symmetry reduction gives metrics on the corresponding
normalized event-orbit spaces, including for both moving frames, while the implementation uses
scalable Hungarian assignment followed by conditional transverse $O(2)$ alignment.
In the matched massless $t\bar t$ comparison, the updated prescription
improves the endpoint AUC over legacy ordering by $4.70\pm0.20$ percentage
points at $N=30$.

Across the complete pipelines, the democratic moving frame has the
largest mean endpoint AUC in all three samples:
$0.9510\pm0.0009$ for $t\bar t$ versus QCD at $N=30$,
$0.8579 \pm 0.0013$ for four-top versus merged $t\bar t+\text{jets}$ at $N=45$,
and $0.9112\pm0.0039$ for conditioned BlackMax B1 versus QCD at $N=30$.
The pivoted frame is second at each endpoint, with the BlackMax difference
less well resolved by the split variation.  The democratic construction
provides a smooth, permutation-equivariant identification of
momentum-closure subspaces, with its linear weight by energy fractions that may give subleading prongs greater influence.  This interpretation alone does not
explain the classification ranking.

More generally, the construction provides a direct route from pseudojet
four-vectors to event classification.  Given a labeled training sample,
the normalization, representation, assignment, and classifier require no
particle identities, hypothesized decay chain, or process-specific feature
engineering.  Exclusive $k_T$ is one way to supply these inputs; other jet
definitions can use the same framework at a common supported multiplicity.
The four-vectors are the only kinematic inputs to the classifier, allowing
the prescription to be transferred between event definitions and signal
hypotheses. 

The method does not require very high values of $N$ to be effective, and zero padding was not
studied.  Finer resolution exposes additional event structure, but its
classification benefit depends on the sample and representation.
The democratic $t\bar t$ branch already reaches a mean AUC of $0.9397$ at
$N=7$ and improves further through $N=30$; four-top and conditioned BlackMax
instead exhibit broad plateaus near $N=15$ and $N=20$--$30$, respectively.
The useful multiplicity should therefore be chosen within the range
supported by the sample, balancing validation performance and computational
cost. 

A central next question is which geometric differences between signal and
background populations produce the discrimination.  Relating their
separation to energy sharing, angular structure, and reconstructed masses
could give the classifier a more direct physical interpretation.
Cross-process AUC differences also reflect the selections: the BlackMax
benchmark deliberately suppresses several otherwise powerful global
discriminants, so its AUC cannot be compared with the $t\bar t$ value as
a measure of intrinsic process similarity. 

A more speculative direction is to tailor the representation to known features of particular processes, analogous to the use of physically
motivated reference geometries in optimal-transport
observables~\cite{Ba2023SHAPER,CesarottiLeBlanc2025}. The labeled collider space is homeomorphic to a closed $(4N-4)$-dimensional ball, leaving considerable freedom in the choice of representation and comparison metric. The spherical-sector and convex spectrahedral representations illustrate how the same kinematic information can support different notions of event similarity. Even with a fixed metric on the target space, different representation maps can induce different event distances and need not yield the same classification performance. Knowledge of kinematic constraints or of how signal and background populate phase space could guide further representations that emphasize features relevant for discrimination. Such constructions could incorporate process information as optional guidance for the geometry, extending the framework beyond the choices studied here, which require neither particle identities nor assumed decay chains.

Improved minimization is a complementary direction.  Alternating
permutation assignment with transverse $O(2)$ alignment, possibly from
several initializations, could improve the joint minimum approximation
while retaining scalable individual steps.  Repeating these steps for
every event pair nevertheless increases the computational cost.
Controlled mass ablations remain useful for separating mass information
from frame and assignment effects; here finite resources were devoted to
comparing substantially different representations and assignment
prescriptions.  Multiclass SVMs and comparisons involving several signal
and background processes would extend the same event distances
beyond the binary benchmarks studied here.
\acknowledgments

We thank Liam Brennan and Danyi Zhang for discussions of high-multiplicity collider samples; Tamas Almos Vami for these discussions as well as a careful reading of the manuscript and helpful comments; and Yu-Tse (Alan) Lee for discussions of geometry. We used computational facilities purchased with funds from the
National Science Foundation (CNS-1725797) and
administered by the Center for Scientific Computing.  The Center for
Scientific Computing is supported by the California NanoSystems Institute
and the Materials Research Science and Engineering Center (NSF DMR 2308708)
at the University of California, Santa Barbara. The authors used OpenAI's ChatGPT and Codex to assist with proofreading
the draft for correctness, consistency, and typographical errors,
refining prose, making minor equation rearrangements, generating and
editing \LaTeX{} code for diagrams, tables, and figure layouts, and
preparing job-submission scripts and running existing analysis code on
the computing cluster. The authors take full responsibility for the
final manuscript and its results.

\appendix
\newpage
\section{Numerical results and comparison details}
\label{app:numerical-validation}

\subsection{Endpoint AUCs and split counts}
\label{app:numerical-status}
\label{app:classifier-protocol}

Table~\ref{tab:endpoint-means} gives the mean test AUC and outer-split
standard error at the endpoint of each complete-pipeline scan.

\begin{table}[H]
\centering
\small
\begin{tabular}{@{}lccc@{}}
\toprule
Branch & $t\bar t$, $N=30$ & Four-top, $N=45$ & BlackMax, $N=30$ \\
\midrule
Pivoted moving frame
 & $0.9392\pm0.0010$ & $0.8500\pm0.0017$ & $0.9057\pm0.0028$ \\
Democratic moving frame
 & $0.9510\pm0.0009$ & $0.8579\pm0.0013$ & $0.9112\pm0.0039$ \\
Radial spherical
 & $0.9314\pm0.0013$ & $0.8207\pm0.0015$ & $0.8832\pm0.0027$ \\
Convex energy--momentum
 & $0.9137\pm0.0015$ & $0.7953\pm0.0016$ & $0.8713\pm0.0019$ \\
Massless lightcone--Hungarian
 & $0.9091\pm0.0015$ & $0.8023\pm0.0024$ & $0.8589\pm0.0023$ \\
\bottomrule
\end{tabular}
\caption{Mean test AUC and outer-split standard error at the endpoint of the
three classification scans.}
\label{tab:endpoint-means}
\end{table}

Table~\ref{tab:ttbar-assignment-endpoint-means} gives the corresponding
assignment-control endpoints.  At matched $c=1$, the Hungarian branch
improves over legacy ordering by $0.04699\pm0.00203$ across their 12 common
outer splits.

\begin{table}[H]
\centering
\small
\begin{tabular}{@{}lcc@{}}
\toprule
Assignment branch & Number of outer splits & Test AUC at $N=30$ \\
\midrule
Legacy lightcone-ordering, $c=0.25$ & $10$ & $0.8655\pm0.0008$ \\
Legacy lightcone-ordering, $c=1$ & $12$ & $0.8642\pm0.0007$ \\
Legacy lightcone-ordering, $c=4$ & $12$ & $0.8625\pm0.0008$ \\
Massless lightcone--Hungarian, $c=1$ & $12$ & $0.9112\pm0.0021$ \\
\bottomrule
\end{tabular}
\caption{Mean test AUC and outer-split standard error at the $t\bar t$
assignment-scan endpoint.}
\label{tab:ttbar-assignment-endpoint-means}
\end{table}

\begin{table}[H]
\centering
\small
\begin{tabular}{@{}lccc@{}}
\toprule
Democratic minus & $t\bar t$, $N=30$ & Four-top, $N=45$ & BlackMax, $N=30$ \\
\midrule
Pivoted
 & $0.01185\pm0.00059$ & $0.00796\pm0.00042$ & $0.00548\pm0.00294$ \\
Convex energy--momentum
 & $0.03734\pm0.00081$ & $0.06267\pm0.00133$ & $0.03994\pm0.00234$ \\
Massless lightcone--Hungarian
 & $0.04188\pm0.00108$ & $0.05567\pm0.00154$ & $0.05230\pm0.00395$ \\
\bottomrule
\end{tabular}
\caption{Paired mean endpoint AUC differences and standard errors across
common outer splits.}
\label{tab:paired-endpoint-differences}
\end{table}

\subsection{Residual scalar separation in BlackMax}
\label{app:blackmax-numerics}

Table~\ref{tab:blackmax-scalars} reports direct rank-based one-variable AUCs
on the full conditioned sample before class balancing, with score orientation
chosen separately for each variable to give an AUC of at least $1/2$.
Before the $M_{30},S_{30}$ conditioning, but after the common $H_T$ window,
acceptance, and multiplicity requirements of
Section~\ref{subsec:blackmax-sample}, the values for
$E_{\mathrm{vis}}^{\mathrm{lab}}$, $M_{30}$, and $S_{30}$ are
$0.9641$, $0.9591$, and $0.9517$, respectively.
The conditioning therefore substantially reduces their separation.
The laboratory-frame energy effective multiplicity is
\begin{equation}
 N_{\mathrm{eff}}^{\mathrm{lab}}
 =\frac{\bigl(\sum_iE_i^{\mathrm{lab}}\bigr)^2}
 {\sum_i(E_i^{\mathrm{lab}})^2},
 \label{eq:blackmax-lab-effective-multiplicity}
\end{equation}
evaluated on the $N=30$ blocks before the center-of-momentum boost.

\begin{table}[H]
\centering
\small
\begin{tabular}{@{}lc@{}}
\toprule
Single variable & Direct AUC after conditioning \\
\midrule
$E_{\mathrm{vis}}^{\mathrm{lab}}=\sum_iE_i^{\mathrm{lab}}$ at $N=30$ & $0.7701$ \\
$S_{30}$ & $0.7586$ \\
$M_{30}$ & $0.7471$ \\
Energy fraction carried by the five leading blocks & $0.7090$ \\
Laboratory-frame $N_{\mathrm{eff}}^{\mathrm{lab}}$ & $0.7062$ \\
Particle-level $H_T$ & $0.5429$ \\
Sum of the $N=30$ block masses & $0.5279$ \\
\bottomrule
\end{tabular}
\caption{Residual one-variable separation in the conditioned BlackMax
B1--QCD sample.  These values use each scalar directly as a score and
provide one-variable reference points.}
\label{tab:blackmax-scalars}
\end{table}

\subsection{Assignment performance}
\label{app:assignment}
\begin{figure}[t]
\centering
\begin{minipage}[c]{0.485\linewidth}
  \centering
  \includegraphics[width=\linewidth]
    {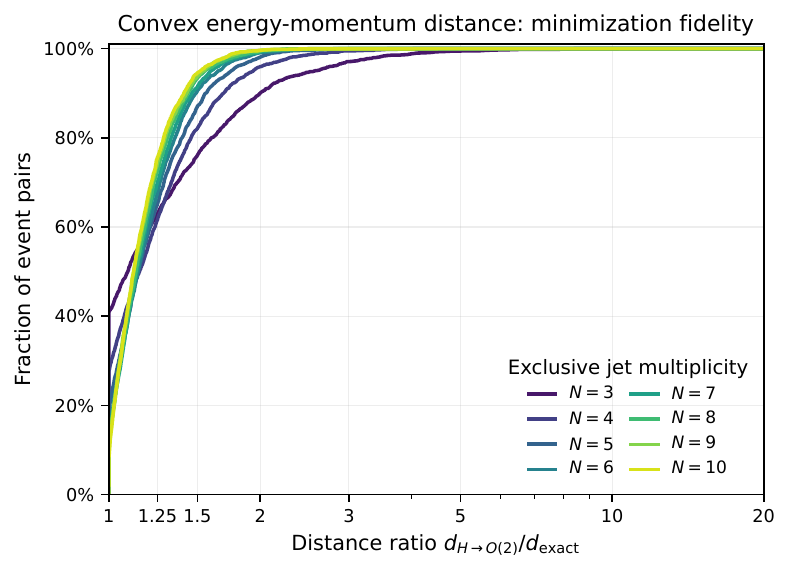}
\end{minipage}\hfill
\begin{minipage}[c]{0.485\linewidth}
  \centering
  \includegraphics[width=\linewidth]
    {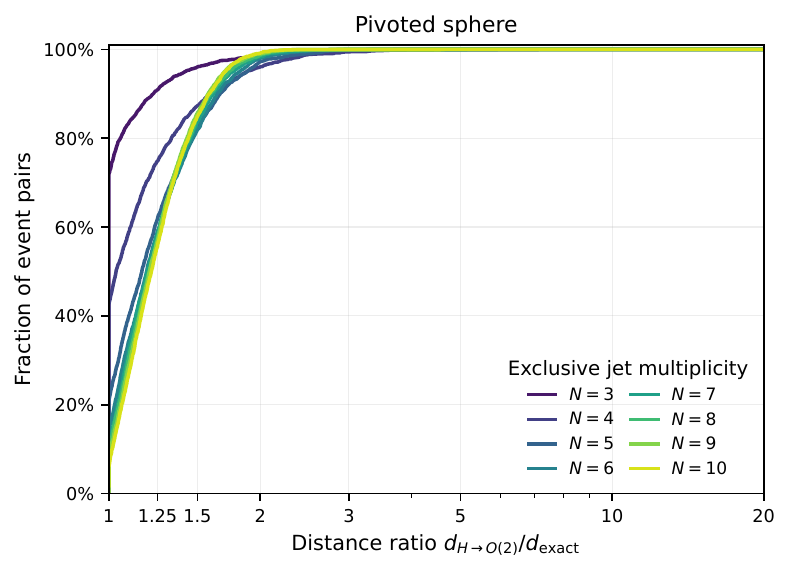}
\end{minipage}

\medskip

\begin{minipage}[c]{0.485\linewidth}
  \centering
  \includegraphics[width=\linewidth]
    {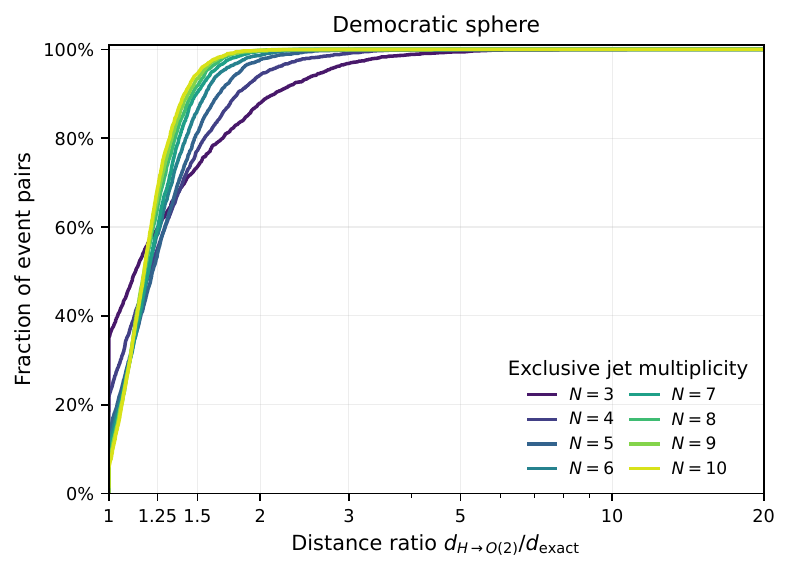}
\end{minipage}\hfill
\begin{minipage}[c]{0.485\linewidth}
  \centering
  \includegraphics[width=\linewidth]
    {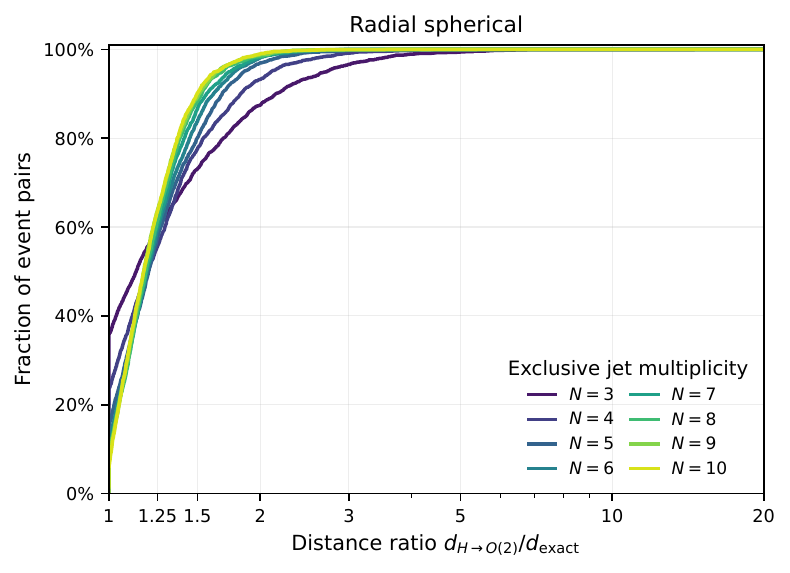}
\end{minipage}
\caption{Empirical cumulative distributions of the distance ratio
for the four mass-aware branches. Colors indicate the exclusive
multiplicity $N$.}
\label{fig:hungarian-distance-ratios}
\end{figure}

\begin{figure}[t]
\centering
\begin{minipage}[c]{0.485\linewidth}
  \centering
  \includegraphics[width=\linewidth]
    {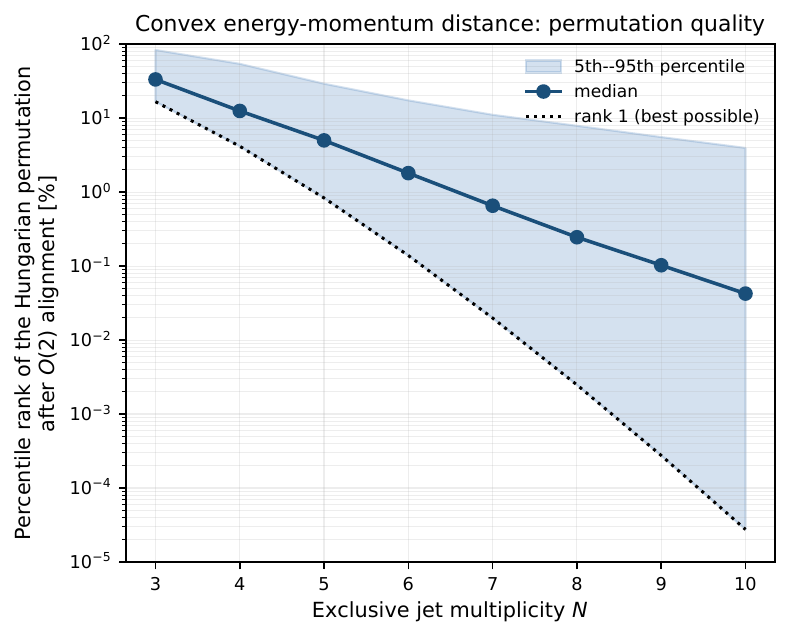}
\end{minipage}\hfill
\begin{minipage}[c]{0.485\linewidth}
  \centering
  \includegraphics[width=\linewidth]
    {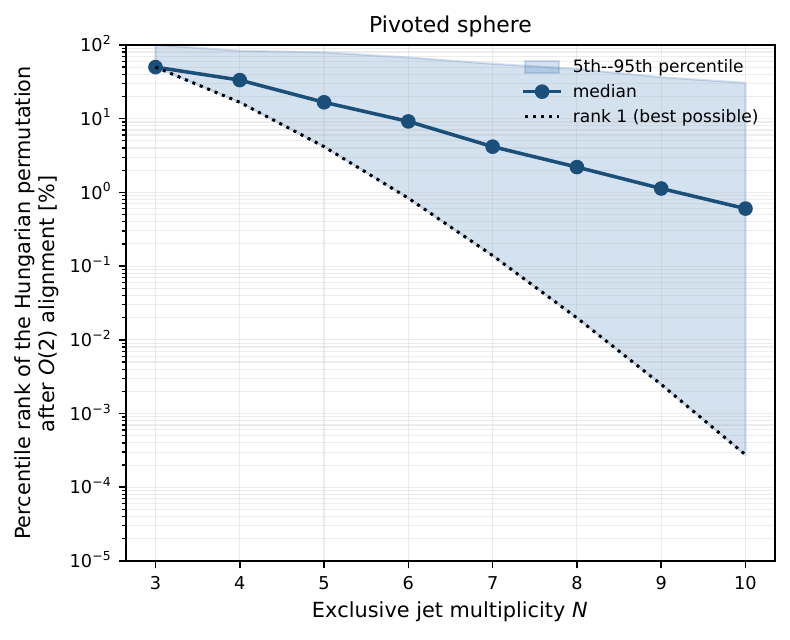}
\end{minipage}

\medskip

\begin{minipage}[c]{0.485\linewidth}
  \centering
  \includegraphics[width=\linewidth]
    {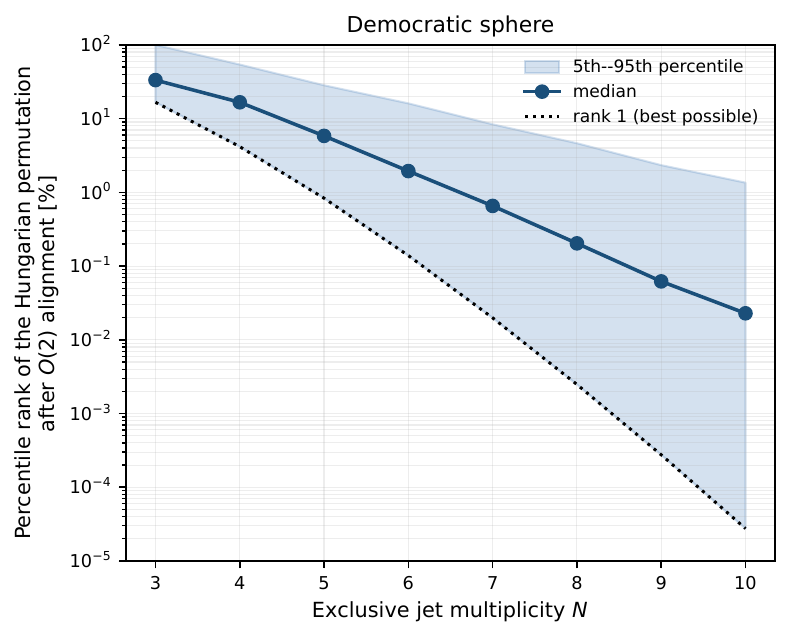}
\end{minipage}\hfill
\begin{minipage}[c]{0.485\linewidth}
  \centering
  \includegraphics[width=\linewidth]
    {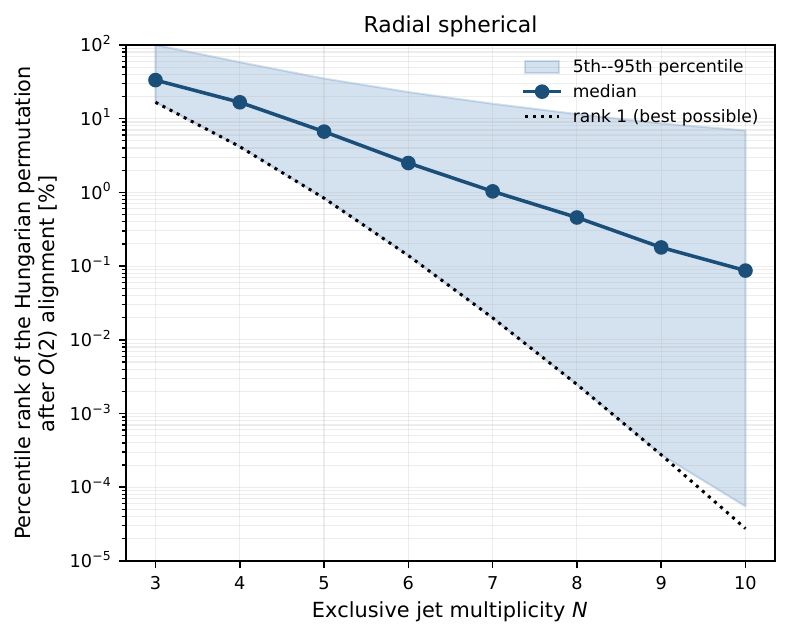}
\end{minipage}
\caption{Permutation-rank percentiles for the same comparisons.
Points show medians, bands the 5th--95th percentile range,
and dotted curves the rank-one floor $100/|G|$.}
\label{fig:hungarian-percentile-ranks}
\end{figure}

We compare all four Section~\ref{sec:collider-metrics} branches on the same $2{,}500$ pairs of normalized $t\bar t$ events at each
$N\leq10$. The reference enumerates every allowed
permutation, minimizing over both sectors of $O(2)$ separately
for each. There are $|G|=(N-1)!$ candidates for the pivoted branch
and $N!$ otherwise.
Fig.~\ref{fig:hungarian-distance-ratios} shows the distance
ratio $D_{H\to O(2)}/D_{\mathrm{joint}}$;
Fig.~\ref{fig:hungarian-percentile-ranks} shows $100r_H/|G|$,
where $r_H$ is the selected permutation's rank after alignment,
with rank one best. At $N = 10$, the median distance ratios range from 1.12 to $1.22$ across the four branches,
with appreciable upper tails. The median rank percentile decreases with
multiplicity in every branch and lies within the best $1\%$ of the allowed
permutations at $N=10$. The democratic and radial branches have similar median
ratios despite different permutation-rank percentiles. These diagnostics
provide complementary measures of assignment quality: the percentile ranks
describe relative position within the factorially growing candidate set,
while the distance ratios quantify the cost of the two-stage approximation.

Exact Hungarian assignment ties are resolved according to solver ordering.
On tie loci, event-exchange symmetry,
$D_{H\to O(2)}(A,B)=D_{H\to O(2)}(B,A)$, requires a tie rule
compatible with exchanging the events.

\subsection{Computational performance}
\label{app:computational-performance}

For $n_{\mathrm{ev}}$ events with $N$ pseudojets each, a dense
$n_{\mathrm{ev}}\times n_{\mathrm{ev}}$ distance matrix requires
$\mathcal O(n_{\mathrm{ev}}^2)$ pair comparisons and matrix storage.
Hungarian assignment gives a worst-case operation count of
$\mathcal O(n_{\mathrm{ev}}^2N^3)$.  The implementation prepares each
event's features once and distributes the pair comparisons across CPU
workers.  Benchmarks of the democratic spherical branch approach the
expected quadratic dependence on event count: at $N=45$, the increases
from $4096$ to $8192$ and then $16384$ events multiply the runtime by
$3.89$ and $3.95$, respectively.

These benchmarks used 40 physical CPU cores on a node with two 24-core
Intel Xeon Platinum 8268 processors at $2.90$~GHz\footnote{These are high-end processors from 2019.}, one numerical-library
thread per worker, and single-precision inputs and matrix storage. 
A complete $16384\times16384$ matrix at $N=45$ took $7.97$~min. 
At fixed $n_{\mathrm{ev}}=2048$ in the four-top sample, the times were
$2.65$, $4.70$, and $13.76$~s for $N=3$, $30$, and $60$, respectively.
This measured multiplicity dependence is substantially milder than the
worst-case bound over the tested range.  These component times are
medians of three fresh computations from preprocessed events in memory,
including feature preparation and matrix construction but excluding
worker startup, persistence, and classification.

On the same hardware configuration, a scan of 8000 $t\bar t$/QCD events
constructed and stored all 16 matrices at $N=3,\ldots,15,20,25,30$ in
$13.54$~min, including distance diagnostics. The full analysis from
preprocessed events, including classification for one outer split and
kernel eigenvalue diagnostics, took $44.25$~min (with optional diagnostics dominating the time).  These wall times demonstrate that
classification studies over a broad multiplicity range are practical
on a single multicore compute node.

\section{The LIPS measure}
\label{app:fixed-lips}
\label{subsec:lips-weighted-area}

This appendix expresses the Lorentz-invariant phase-space measure
in the radial coordinates of Section~\ref{subsec:fixed-mass-shell}
and explains the inclusion of zero-momentum configurations.
Throughout, $N\geq2$, $m_i\geq0$, and $M>\sum_i m_i$.

\subsection{Soft particles}

For total momentum $P=(M,\bm0)$, define
$\delta_+(p^2-m^2)=\theta(p^0)\delta(p^2-m^2)$ and use
\begin{equation}
 \dd\mathrm{LIPS}_N(P;\bm m)
 =(2\pi)^{4-3N}
 \left[\prod_{i=1}^N
   \dd^4p_i\,\delta_+(p_i^2-m_i^2)\right]
 \delta^{(4)}\!\left(P-\sum_i p_i\right).
 \label{eq:full-lips}
\end{equation}
Integrating over each $p_i^0$ leaves the on-shell factor
$\dd^3\bm p_i/(2E_i)$.
For a massless particle, $E=\norm{\bm p}$, so
\begin{equation}
 \frac{\dd^3\bm p}{2\norm{\bm p}}
 =\frac12 E\,\dd E\,\dd\Omega,
 \qquad
 \int_{0<\norm{\bm p}<\epsilon}
 \frac{\dd^3\bm p}{2\norm{\bm p}}
 =\pi\epsilon^2.
 \label{eq:massless-one-particle-measure}
\end{equation}
The apparent Cartesian singularity at zero momentum is therefore
locally integrable.

The same conclusion holds after imposing total four-momentum
conservation, including simultaneous soft limits. Let $S$ be the
set of all $k$ zero-energy legs at a given configuration, and $H$
its complement. Writing
$q_a=(\norm{\bm q_a},\bm q_a)$ and $Q=\sum_{a\in S}q_a$, the measure
factorizes as
\begin{equation}
     \dd\mathrm{LIPS}_N(P;\bm m)
 =
 \left[\prod_{a\in S}
   \frac{\dd^3\bm q_a}
        {2(2\pi)^3\norm{\bm q_a}}\right]
 \dd\mathrm{LIPS}_{N-k}(P-Q;\bm m_H).
\end{equation}
To obtain this, we repeatedly used the two-body recursive formula~\cite{PDGKinematics2025} and integrated out the intermediate coordinates.\footnote{This can also be obtained directly by factorizing the measure.} At $Q=0$, the remaining subsystem is above threshold, and at least
two of its particle velocities differ: equal velocities would
force all spatial momenta to vanish in the COM frame, contradicting
$M>\sum_i m_i$. The four conservation constraints therefore have
full rank in the remaining momenta, so the remaining measure has
a continuous, strictly positive density in suitable local
coordinates for small $\bm q_a$.

Each soft factor is proportional to
$E_a\,\dd E_a\,\dd\Omega_a$, whose integral over
$0<E_a<\epsilon$ is positive and of order $\epsilon^2$.
\textbf{Their product is locally integrable}, assigns zero measure to the
soft strata, and gives positive measure to every neighborhood
of each such configuration. These configurations therefore
belong to the support of LIPS. Retaining them in the closed
phase space of Eq.~\eqref{eq:phase-space-body} leaves ordinary
LIPS integrals unchanged. Amplitudes or measurement functions
may still diverge on approach; such divergences concern the
integrand rather than a delta-atom at zero momentum.

\subsection{The measure in radial coordinates}

An orthogonal Jacobi transformation separates the COM coordinate
from the $3N-3$ relative momentum coordinates, with unit Jacobian.
Since
\[
 \bm\kappa_0=\frac{1}{\sqrt N}\sum_i\bm p_i,
 \qquad
 \delta^{(3)}\!\left(\sum_i\bm p_i\right)
 =N^{-3/2}\delta^{(3)}(\bm\kappa_0),
\]
integrating the momentum-conservation delta function leaves
$N^{-3/2}$ times Euclidean volume on the momentum-conserving
subspace. Writing $\bm p=ru$ with $u\in\hat S$, as in
Section~\ref{subsec:fixed-mass-shell}, this volume is
$r^{3N-4}\,\dd r\,\dd\Omega_{3N-4}(u)$, where
$\dd\Omega_{3N-4}$ is the round area measure on the unit sphere.

With
\[
 E_i(r,u)=\sqrt{m_i^2+r^2\norm{\bm u_i}^2},
 \qquad
 g_u(r)=\sum_i E_i(r,u),
\]
Eq.~\eqref{eq:full-lips} becomes
\begin{equation}
 \dd\mathrm{LIPS}_N
 =
 \frac{(2\pi)^{4-3N}}{2^N N^{3/2}}\,
 \frac{r^{3N-4}\,\dd r\,\dd\Omega_{3N-4}(u)}
      {\prod_i E_i(r,u)}
 \delta\!\left(g_u(r)-M\right).
 \label{eq:prepolar-shell-measure}
\end{equation}
The unique positive solution $r_*(u)$ of $g_u(r)=M$ is the
radial function defined in Eq.~\eqref{eq:radius-equation}.
On the soft-free locus, where all $E_i>0$,
\[
 g_u'(r)
 =r\sum_i\frac{\norm{\bm u_i}^2}{E_i(r,u)}>0,
 \qquad r>0.
\]
Integrating the remaining delta function therefore gives
\begin{equation}
 \dd\mathrm{LIPS}_N
 =
 \frac{(2\pi)^{4-3N}}{2^N N^{3/2}}\,
 \frac{r_*^{\,3N-5}}{\prod_i E_i}
 \left[
   \sum_i\frac{\norm{\bm u_i}^2}{E_i}
 \right]^{-1}
 \dd\Omega_{3N-4}(u),
 \label{eq:fixed-shell-measure-sphere}
\end{equation}
where $r_*=r_*(u)$ and $E_i=E_i(r_*(u),u)$.
This expresses LIPS as a generally nonuniform weight times
the round sphere measure. The formula holds almost everywhere;
at zero-energy configurations, the measure is understood
through the locally integrable extension established above.

For $N=2$, $\norm{\bm u_1}^2=\norm{\bm u_2}^2=1/2$ and
$r_*=\sqrt2\,p_*$, so Eq.~\eqref{eq:fixed-shell-measure-sphere}
reduces to the standard two-body result
\[
 \dd\mathrm{LIPS}_2
 =\frac{p_*}{16\pi^2 M}\,\dd\Omega_2,
 \qquad
 p_*=\frac{\sqrt{\lambda(M^2,m_1^2,m_2^2)}}{2M},
\]
where
$\lambda(a,b,c)=a^2+b^2+c^2-2ab-2ac-2bc$
is the Källén function~\cite{PDGKinematics2025}.
\section{Topology, coordinate relations, and exact metrics}
\label{app:metricity}

This appendix supplies the arguments deferred in Sections \ref{sec:fixed-shell} and \ref{sec:collider-metrics}. For the fixed-mass radial map, the main text explains why each momentum direction determines exactly one event; we complete that argument by checking continuity. While continuity is obvious from a physics perspective, for completeness we include a proof. In principle, a single radial solution could jump between momentum configurations. For the variable-mass space, we make the radial rescaling explicit and verify that imposing nonnegative masses gives a closed ball.

The main reconstruction problem arises in the democratic representation. Section \ref{subsec:moving-frame-features} shows that its features lie in the stated spherical sector. Here we state that every point of that sector represents exactly one labeled event and that reconstruction is continuous, including at zero energy. These results support the claim that radial and democratic coordinates retain the same fixed-scale event information while defining different distances. We then treat the restricted pivoted representation and prove the metric statements for exact symmetry reduction.
\subsection{Fixed-mass radial topology}
\label{app:radial-topology}
Fix $N\geq2$, masses $m_i\geq0$, and center-of-mass energy $M>0$. To slow down the argument in the main text, let's define the momentum conserving subspace explicitly
\begin{equation}
 V_N=\left\{(\bm p_1,\ldots,\bm p_N)\in(\R^3)^N:
 \sum_{i=1}^N\bm p_i=0\right\},
 \qquad \dim V_N=3N-3,
 \label{eq:VN}
\end{equation}
and thus phase space is 
\begin{equation}
 \Phi_N(M;\bm m)=
 \left\{\bm p\in V_N:
 \sum_{i=1}^N\sqrt{m_i^2+\norm{\bm p_i}^2}=M\right\}.
 \label{eq:phase-space}
\end{equation}
It is empty for $M<\sum_i m_i$ and is a single point for
$M=\sum_i m_i$, so we assume $M>\sum_i m_i$. 

Equip $V_N$ with the inherited Euclidean inner product
$
 \langle\bm p,\bm q\rangle_{V_N}
 =\sum_i\bm p_i\cdot\bm q_i,
$
 so we can define the unit sphere through its norm 
\[
 \hat{S} = S(V_N)=\{u\in V_N:\norm{u}_{V_N}=1\}
\]
The radius $r_*(u)$ is defined by solving the energy constraint $g_{\bm u}(r) = M$. Having exactly one solution in each direction does not, by itself, show that nearby directions give nearby radii. Thus, we need to show continuity. The proof below establishes this: the radii remain bounded, and any limiting radius must solve the limiting energy equation, whose solution is unique.
\begin{theorem}[Sphere topology of fixed-mass phase space]
\label{thm:sphere-topology}
Fix $N\geq2$, $m_i\geq0$, and $M>\sum_i m_i$.  The closed shell in
Eq.~\eqref{eq:phase-space} is homeomorphic to $S(V_N)\cong S^{3N-4}$.
The mutually inverse maps are Eq.~\eqref{eq:radial-homeomorphism}, where
$r_*(u)$ is the unique positive solution of \begin{equation*}
    g_{\bm u}(r)=\sum_{i=1}^N\sqrt{m_i^2+r^2\norm{\bm u_i}^2}=M
\end{equation*}
\end{theorem}

\begin{proof}
For fixed $u$, $g_u$ is continuous and strictly increasing,
$g_{\bm u}(0)=\sum_i m_i<M$, and $g_u(r)\to\infty$.  Thus every ray meets the
shell exactly once.  This unique radial intersection makes $\Psi$ bijective.
To check continuity, let $\bm u_n\to \bm u$.  Since
$g_{\bm u_n}(r_*(\bm u_n))=M$ and
$M\geq r_*(u_n)\sum_i\norm{\bm u_{n,i}}\geq r_*(\bm u_n)$, every subsequence of the
radii has a convergent subsubsequence.  Joint continuity of $g_{\bm u}(r)$ and
uniqueness of its root force every such limit to be $r_*(\bm u)$, so
$r_*(\bm u_n)\to r_*(\bm u)$.
Hence $\Psi$ is continuous.  Its source is compact and its target is
\textit{Hausdorff},\footnote{In this appendix, this word means a subset of Euclidean space with its usual topology.} so the compact-to-Hausdorff closed-map lemma makes $\Psi$ a
homeomorphism~\cite{lee2011topologicalmanifolds}.\footnote{The
same argument applies in \(D\)-dimensional Minkowski spacetime, giving LIPS the topology \(S^{(D-1)N-D}\).  Furthermore, while the argument is most natural in
the center-of-mass frame, the topology holds for any future-timelike total
momentum \(P\).}
\end{proof}
The homeomorphism gives two immediate consequences. The first recovers the spherical topology of massless phase space, discussed in Section~\ref{subsec:spinor-connection-main}, with an explicit expression for the radial function. The second verifies that the angular distance introduced in Eq.~\eqref{eq:fixed-shell-radial-metric} defines a metric compatible with the original phase-space topology. 
\begin{corollary}[Massless shell]
\label{cor:massless-sphere}
For $N\geq2$ and $M>0$, the massless shell has topology $S^{3N-4}$,
with radius given by \begin{equation}
     r_*(\bm u)=\frac{M}{\sum_i\norm{\bm u_i}}
\end{equation}
\end{corollary}

\begin{corollary}[Fixed-mass radial metric]
\label{cor:fixed-shell-radial-metric}
For $N\geq2$, $m_i\geq0$ and $M>\sum_i m_i$, the distance
$d_{\Phi,\mathrm{rad}}$ in Eq.~\eqref{eq:fixed-shell-radial-metric}
is a metric inducing the subspace topology on $\Phi_N(M;\bm m)$.
\end{corollary}
\begin{proof}
The map $\bm p\mapsto\bm p/\norm{\bm p}_{V_N}$ is a homeomorphism
onto $S(V_N)$ by Theorem~\ref{thm:sphere-topology}.  The stated distance
is its pullback of the round angular metric.
\end{proof}
This fixed-mass metric normalizes only the spatial momenta.  For nonzero
masses it need not equal the restriction of the mass--momentum radial
metric in Eq.~\eqref{eq:collider-radial-metric}.

\subsection{Adjoining the fixed-mass shells}
\label{app:adjoining-topology}
Next, we supply the complete proofs for the collider topologies in the body. The proof of the sphere space is already substantially in the text: we are fixing the norm of a set and thus we end up with a sphere. The collider dome  proof is omitted from the main text as this is physically quite simple: we start by restricting a single mass coordinate to $m_N\geq 0$, meaning we are cutting a sphere in two hemispheres, and discarding one of them. The remaining restrictions $m_i\geq0$ then select a convex portion of this ball with nonempty interior, which is itself topologically a closed ball. The proof below makes this argument precise and verifies that our low-dimensional \textit{projection} intuition carries over.

To distinguish a variable signed coordinate from the fixed mass parameters
$m_i$ of $\Phi_N(M;\bm m)$, write $s_i$ for the mass coordinate in this
appendix.  On the physical sector, $s_i\geq0$ is the coordinate denoted
$m_i$ in the main text; a fixed-mass slice imposes $s_i=m_i$.
Define
\begin{equation}
 W_N=V_N\oplus\R_s^N
 =\left\{((\bm p_i,s_i))_{i=1}^N:\sum_i\bm p_i=0\right\},
 \qquad \dim W_N=4N-3,
 \label{eq:collider-ambient-space}
\end{equation}
with
\[
 E_i=\sqrt{\norm{\bm p_i}^2+s_i^2},\qquad
 \norm{\mathsf w}_{W_N}^2
 =\sum_i(\norm{\bm p_i}^2+s_i^2)=\sum_iE_i^2,\qquad
 \norm{\mathsf w}_{E}=\sum_i E_i.
\]
Here $\norm{\cdot}_{W_N}$ is the Euclidean norm denoted
$\norm{\cdot}$ in Eq.~\eqref{eq:collider-block-norm}, whereas
$\norm{\cdot}_{E}$ is the sum of the Euclidean block norms and is
itself a norm on $W_N$.
Write $S(W_N)=\{u\in W_N:\norm{u}_{W_N}=1\}$.  In this notation,
\[
 \cC_N^{\mathrm{sgn}}(M)=\{\mathsf w\in W_N:\norm{\mathsf w}_E=M\},
 \qquad
 \cC_N^+(M)=\{\mathsf w\in\cC_N^{\mathrm{sgn}}(M):s_i\geq0\}.
\]
Note that this enlarged space includes the zero-momentum configuration $\sum_i s_i = M$, but indeed the unphysical sector $\sum_i s_i >M $ is automatically excluded. In general, the masses and the momenta cannot be chosen independently in this space which motivates a careful proof of the topologies. 

\begin{theorem}[Topology of the collider dome and signed space.]
\label{thm:collider-topology}
For $M>0$ and $N\geq2$, the signed shell $\cC_N^{\mathrm{sgn}}(M)$ is
homeomorphic to $S^{4N-4}$ and its nonnegative-mass sector $\cC_N^+(M)$
is homeomorphic to the closed ball $B^{4N-4}$.
\end{theorem}

\begin{proof}
For $u=(u_1,\ldots,u_N)\in S(W_N)$, with $u_i\in\R^4$, set
\begin{equation}
 \mathcal S(u)=\sum_i\norm{u_i}_{\R^4},
 \qquad r_{\mathrm{blk}}(u)=\frac{M}{\mathcal S(u)}.
\end{equation}
Then $u\mapsto r_{\mathrm{blk}}(u)u$ is a homeomorphism from $S(W_N)$ to
$\cC_N^{\mathrm{sgn}}(M)$, with inverse
$\mathsf w\mapsto\mathsf w/\norm{\mathsf w}_{W_N}$.  This proves
the sphere statement.

Radial normalization identifies the physical sector with
\begin{equation}
 S(V_N\oplus\R^N)\cap\{s_i\geq0\}.
 \label{eq:spherical-orthant}
\end{equation}
Project this spherical sector onto
$(\bm p,s_1,\ldots,s_{N-1})$. Its image is
\[
 \left\{(\bm p,s_1,\ldots,s_{N-1})
 \in V_N\oplus\R^{N-1}:
 \norm{\bm p}^2+\sum_{i=1}^{N-1}s_i^2\leq1,
 \quad s_i\geq0\right\}.
\]
The projection is a homeomorphism, since the omitted coordinate
is recovered continuously as
\[
 s_N=\sqrt{1-\norm{\bm p}^2-\sum_{i=1}^{N-1}s_i^2}.
\]
The image is a compact convex body with nonempty interior in
the $(4N-4)$-dimensional space $V_N\oplus\R^{N-1}$,
and is therefore homeomorphic to $B^{4N-4}$.
\end{proof}

The spherical orthant's massless faces and zero-block loci correspond
to the lower-rank strata in the spinor description of
Appendix~\ref{app:gram-geometry}.
\subsection{The democratic representation on the closed collider space}
\label{app:dem-closed-homeomorphism}

The democratic representation in Section \ref{subsec:moving-frame-features} was introduced to change the relative weighting of energetic and subleading objects while retaining the event’s kinematic information. The equations there show that every physical event maps continuously into a fixed spherical sector. What remains is to show that this sector is filled completely, that no two different labeled events have the same features, and that reconstruction from the features is continuous. These results allow us to take the metric on the unit sphere and map it back to the collider space and invert it, defining a distance directly on the collider dome.

Fix $N\geq2$ and $M>0$.  Write an event as
$\mathsf w=((\bm p_i,s_i))_{i=1}^N$, with
$E_i=(\norm{\bm p_i}^2+s_i^2)^{1/2}$.  The signed and physical shells are
\begin{align}
 \cC_N^{\mathrm{sgn}}(M)
 &=\left\{\mathsf w\in(\R^3\times\R)^N:
       \sum_i\bm p_i=0,\quad\sum_i E_i=M\right\},
 \label{eq:dem-closed-domain}\\
 \cC_N^+(M)
 &=\{\mathsf w\in\cC_N^{\mathrm{sgn}}(M):s_i\geq0\text{ for all }i\}.
 \nonumber
\end{align}
Both include zero-energy blocks.  Such a block necessarily has
$\bm p_i=0$ and $s_i=0$.

Set $r=e_{\mathrm{dem}}=N^{-1/2}(1,\ldots,1)^T$ and define
\begin{equation}
 \nu_i=\sqrt{E_i/M},\qquad
 (\bm q_i,\mu_i)=
 \begin{cases}
   (\bm p_i,s_i)/\sqrt{ME_i},&E_i>0,\\
   (0,0),&E_i=0.
 \end{cases}
 \label{eq:dem-closed-chart}
\end{equation}
Let $Q$ be the $N\times3$ matrix with rows $\bm q_i^T$.  These variables obey
\begin{equation}
 \norm{\nu}^2=1,\qquad
 \nu_i^2=\norm{\bm q_i}^2+\mu_i^2,\qquad Q^T\nu=0.
 \label{eq:dem-closed-identities}
\end{equation}
In matrix notation, the democratic Householder reflection and feature map are
\begin{equation}
 H_r(\nu)=I-\frac{(\nu+r)(\nu+r)^T}{1+r\cdot\nu},\qquad
 \Theta_{\mathrm{dem}}(\mathsf w)=(H_r(\nu)Q,\mu).
 \label{eq:dem-closed-map}
\end{equation}
Writing the three transformed spatial columns as a matrix is only a
reordering of the coordinates in the democratic feature vector.
Define its reference space and physical spherical sector by
\begin{align}
 \mathcal H_{\mathrm{dem}}
 &=\{(A,\mu)\in\R^{N\times3}\times\R^N:A^Tr=0\},
 \nonumber\\
 \Sigma_{\mathrm{dem}}
 &=\{(A,\mu)\in\mathcal H_{\mathrm{dem}}:
       \norm{A}_F^2+\norm{\mu}^2=1,\quad \mu_i\geq0\text{ for all }i\}.
 \label{eq:dem-closed-target}
\end{align}
Here $\mu=(\mu_i)_{i=1}^N$ is the normalized mass vector defined in
Eq.~\eqref{eq:dem-closed-chart}, and $\norm{A}_F$ denotes the Frobenius norm, which is the root of the sum of the squares of the entries of a matrix \begin{equation}
    \norm{A}_F = \sqrt{\sum_{i=1}^{N}\sum_{j=1}^{3} A_{ij}^2}
\end{equation}
The formal argument for the democratic homeomorphism is very technical and uninteresting.   The central challenge with surjectivity is the recovery of the energy fraction square roots $\nu$ when starting from the targets. Concretely, given an $(A,\mu)\in S(\mathcal H_{\mathrm{dem}})$ or $\Sigma_\text{dem}$, to recover $(\bm p,s)$, one needs the Householder involution $H_r(\nu)$ which requires the energy fractions to be built. With it, $H_r(\nu)A$ and $\mu$ are sufficient to recover the inverse coordinates. To show it, one can use Brouwer’s fixed point theorem~\cite{lee2011topologicalmanifolds} on a simplex of trial energy fractions in $\mathcal{H}_\text{dem}$. Injectivity can be shown by contradiction: assume that two distinct events produce the same features. Their mass-shell conditions and momentum conservation yield a positivity identity that forces their energy fractions to agree. The inverse reflection then gives identical momenta and masses, so the two events must coincide.\footnote{The correspondence is indeed a homeomorphism: the forward map is continuous, including at zero energy, and compactness of the closed energy shell ensures continuity of the inverse.} Injectivity in particular is crucial to take our metric on the unit sphere and define a metric on our collider dome.

For the pivoted representation, the same reconstruction argument applies after restoring the omitted zero spatial row and replacing the democratic reference by $r=e_N$. Injectivity follows from the same positivity argument, with the assumption $E_i>0$ keeping the relevant inequalities strict. The pivot variant can only be shown to be injective in the restricted pivot domain which is sufficient for metricity. We will thus use the following statements without proof.

At fixed $M>0$ and $N$, the democratic representation defines homeomorphisms
\begin{equation}
    \Theta_{\mathrm{dem}}:
\mathcal C_N^{\mathrm{sgn}}(M)\longrightarrow S(\mathcal H_{\mathrm{dem}}),
\qquad
\Theta_{\mathrm{dem}}:
\mathcal C_N^+(M)\longrightarrow\Sigma_{\mathrm{dem}}.
\end{equation}
Thus every point of the closed physical spherical sector represents exactly one labeled event, and both the feature map and its inverse are continuous, including at zero energy.

For the pivoted representation, let $X_{\mathrm{piv}}$ consist of positive-energy events whose unique highest-$p_T$ object occupies slot $N$. The map
\begin{equation}
    \Theta_{\mathrm{piv}}:X_{\mathrm{piv}}\longrightarrow S^{4N-4}
\end{equation}
is injective and is the restriction of a smooth map on the positive-energy domain. It is equivariant under $S_{N-1}\times O(2)$: permuting the non-pivot objects or applying a common transverse rotation or reflection produces the corresponding transformation of the features. 

These statements allow us to establish metricity of the associated distances. The event comparisons in the main text also account for the allowed particle permutations and transverse rotations and reflections. We next show that exact minimization over these symmetries gives a metric on the corresponding orbit space. 

\subsection{Metrics under exact symmetry reduction}
\label{app:exact-metrics}
The importance of this section is to show that the symmetry minimizations do not destroy the metric properties. We first show a general statement covering any compact group and apply it to our particular metrics. 

For any map $F:X\to(Y,d_Y)$, the pullback
$d_F(x,x')=d_Y(F(x),F(x'))$ is a pseudometric, and is a metric precisely
when $F$ is injective.  If $F$ is a homeomorphism onto its image, this
metric also induces the original topology of $X$.  These elementary
observations and the following lemma give the metric statements for all
four mass-aware representations.

Let $[x]$ denote the \textit{orbit} of $x$ with respect to our symmetry group. This is the set of events obtained from $x$ by the allowed relabelings and transverse
transformations.

\begin{lemma}[Exact compact-group quotient]
\label{lem:compact-quotient-metric}
Let $(X,d)$ be a metric space and let a compact group $\Gamma$ act
continuously on $X$ by isometries.  Then
\begin{equation}
 \bar d([x],[y])=\min_{\gamma\in\Gamma}d(x,\gamma y)
 \label{eq:compact-quotient-metric}
\end{equation}
is a metric inducing the quotient topology on $X/\Gamma$.
\end{lemma}

\begin{proof}
Compactness and continuity imply that the minimum is attained.  Isometric
invariance makes the definition independent of the representatives, and
replacing $\gamma$ by $\gamma^{-1}$ gives the symmetry axiom.  If
$\bar d([x],[y])=0$, attainment gives $x=\gamma y$ for some $\gamma$, so
$[x]=[y]$.  If $\gamma_{xy}$ and $\gamma_{yz}$ are minimizers, then
\begin{align}
 \bar d([x],[z])
 &\leq d(x,\gamma_{xy}\gamma_{yz}z)\nonumber\\
 &\leq d(x,\gamma_{xy}y)
 +d(\gamma_{xy}y,\gamma_{xy}\gamma_{yz}z)\nonumber\\
 &=\bar d([x],[y])+\bar d([y],[z]),
\end{align}
which proves the triangle inequality.  For the orbit projection
$q:X\to X/\Gamma$,
\[
 q(B_d(x,\epsilon))=B_{\bar d}([x],\epsilon),\qquad
 q^{-1}(B_{\bar d}([x],\epsilon))
 =\bigcup_{\gamma\in\Gamma}B_d(\gamma x,\epsilon).
\]
Thus $q$ is continuous and open for the metric topology, which is exactly
the quotient topology.  No freeness of the action is required.
\end{proof}

Lastly, we show that after exact assignment and alignment, all our metrics remain metrics. We are simply applying the lemma to our particular metrics here to finalize the formalization.
\begin{corollary}[Representation metrics and exact symmetry reduction]
\label{cor:representation-metrics}
At fixed $M>0$ and $N\geq2$, the radial, democratic, and convex distances
in Eqs.~\eqref{eq:collider-radial-metric},
\eqref{eq:moving-frame-metric}, and
\eqref{eq:labelled-fourvector-euclidean} are metrics inducing the original
topology of the closed physical collider space.  The convex statement
uses the identification $F:\cC_N^+(M)\to\cK_N(M)$.
Let $\Gamma=S_N\times O(2)$. For each of these distances,
\begin{equation}
 D_{\mathrm{joint}}([A],[B])
 =\min_{\gamma\in\Gamma}
 d_{\mathrm{rep}}(A,\gamma\cdot B).
 \label{eq:representation-exact-quotient}
\end{equation}
is a metric inducing the corresponding quotient topology.  The democratic
map also induces a homeomorphism
\begin{equation}
 \cC_N^+(M)/\Gamma\ \longrightarrow\ \Sigma_{\mathrm{dem}}/\Gamma.
 \label{eq:dem-closed-quotient-homeomorphism}
\end{equation}

The pivoted angular distance is a metric on $X_{\mathrm{piv}}$
and on its feature image. Its exact symmetry reduction is a
metric on both orbit spaces for
$\Gamma_{\mathrm{piv}}=S_{N-1}\times O(2)$, with the
distinguished slot held fixed.
All statements also hold with the transverse factor omitted.
\end{corollary}

\begin{proof}
The radial and democratic maps are homeomorphisms onto the same closed
spherical sector.  Pulling back the round angular metric
therefore gives metrics with the stated topology.  The convex map $F$
and its inverse are continuous, and the Euclidean metric on $\cK_N(M)$
induces its subspace topology.  The pivoted metric follows from
the statements in Section~\ref{app:dem-closed-homeomorphism}.

Radial normalization and the convex map commute with relabeling and
transverse orthogonal transformations.  For the democratic reference,
$\Pi r=r$ for every permutation matrix, so
$H_r(\Pi\nu)=\Pi H_r(\nu)\Pi^T$.  Permutations act on the rows of
$Q$ and the components of $\mu,\nu$, whereas transverse transformations act on the spatial columns
of $Q$ and preserve the energies.  The zero-block convention respects
both actions.  The democratic map is therefore equivariant, as is its
inverse.  

For the pivoted representation, every allowed permutation $\Pi$ fixes
slot $N$, so $\Pi e_N=e_N$ and hence
\[
H_{\mathrm{piv}}(\Pi\nu)
=\Pi H_{\mathrm{piv}}(\nu)\Pi^T.
\]
Together with the same transverse equivariance, this establishes
equivariance of the pivoted map under $S_{N-1}\times O(2)$.

The target actions are orthogonal and hence isometric.  The quotient
metrics follow from Lemma~\ref{lem:compact-quotient-metric}.
Equivariance of the democratic homeomorphism and its inverse gives
mutually inverse continuous maps on the orbit spaces, proving
Eq.~\eqref{eq:dem-closed-quotient-homeomorphism}.
\end{proof}

\begin{remark}[Intrinsic distance in convex coordinates]
\label{rem:convex-length-distance}
Every finite-length curve joining $A,B\in\cK_N(M)$ has length
at least their Euclidean distance, while convexity ensures
that the straight segment attains this bound. It is the
unique constant-speed minimizing curve. The same argument
applies to the convex positive-scale cone
$\bigcup_{M>0}\cK_N(M)$.
\end{remark}

These results establish metricity of the exact joint minimum
in Eq.~\eqref{eq:event-joint-target}. They do not establish
metricity of the sequential procedure in
Section~\ref{subsec:exact-vs-production}, whose approximation
performance is assessed in Appendix~\ref{app:assignment}.
\section{Spinor frames and the collider spectrahedron}
\label{app:spinor-frames}
This appendix relates spinor-helicity descriptions to the event spaces introduced in the main text. Its central result identifies the collider spectrahedron, written in matrix form, with a quotient of normalized Gram factors. We also specify the open domain covered by the earlier classifier's simplex--sphere parametrization and explain its relation to the closed massless phase space.  
\subsection{Massless spinor frames and lightcone coordinates}
\label{app:spinor-lightcone}

The spinor construction relates the original massless representation to
the variable-mass event space. Write
$p_i^\mu=(E_i,\bm p_i)$, use signature $(+,-,-,-)$, and associate to each
momentum the Hermitian matrix%
\footnote{With the standard definitions
$\sigma^\mu=(I_2,\bm\sigma)$ and $\bar\sigma^\mu=(I_2,-\bm\sigma)$,
this is $\mathsf P_i=p_{i,\mu}\bar\sigma^\mu$.
We suppress Weyl indices and write the spinor factors in matrix notation.}
\begin{equation}
 \mathsf P_i
 =E_iI_2+\bm p_i\cdot\bm\sigma
 =
 \begin{pmatrix}
 E_i+p_{z,i}&p_{x,i}-ip_{y,i}\\
 p_{x,i}+ip_{y,i}&E_i-p_{z,i}
 \end{pmatrix}.
 \label{eq:spinor-bispinor-main}
\end{equation}
Here $\bm\sigma$ denotes the Pauli matrices and
$\det\mathsf P_i=p_i^2$. For a nonzero future-directed
massless momentum, this matrix is positive semidefinite of rank one and
factors as
\begin{equation}
 \mathsf P_i
 =\lambda_i\widetilde\lambda_i^{\,T}
 =\lambda_i\lambda_i^\dagger,
 \qquad
 \widetilde\lambda_i=\bar\lambda_i.
 \label{eq:massless-factor-main}
\end{equation}
Here $\lambda_i\in\mathbb C^2$ is a column spinor and $\bar\lambda_i$
denotes componentwise complex conjugation on the real momentum slice.
The particlewise little-group action is
$(\lambda_i,\widetilde\lambda_i)\mapsto
(e^{i\varphi_i}\lambda_i,e^{-i\varphi_i}\widetilde\lambda_i)$.
Zero momentum is represented by $\lambda_i=0$.

\begin{proposition}[Massless spinor-frame space]
For $N\geq2$ massless momenta with nonnegative energies and total momentum
$P^\mu=(M,\bm0)$, where $M>0$, the normalized spinor frames form
\begin{equation}
 V_2(\mathbb C^N)
 =\{\Lambda\in\mathbb C^{N\times2}:
   \Lambda^\dagger\Lambda=I_2\}
 \cong U(N)/U(N-2).
 \label{eq:massless-stiefel}
\end{equation}
The physical momentum shell, including zero-momentum blocks, is the
topological orbit space
\begin{equation}
 \Phi_N(M;\bm0)
 \cong V_2(\mathbb C^N)/U(1)^N.
 \label{eq:massless-stiefel-quotient}
\end{equation}
\end{proposition}

\begin{proof}
Collect the spinors into
$L=(\lambda_1,\ldots,\lambda_N)\in\mathbb C^{2\times N}$.
Momentum conservation is
$LL^\dagger=MI_2$. Hence
$\Lambda=L^\dagger/\sqrt M$ has two orthonormal columns and lies in
$V_2(\mathbb C^N)$; conversely, every such frame defines a conserved
massless configuration. The momentum map is unchanged by independent row
phases. Two frames give the same positive-semidefinite matrices of rank at
most one precisely when they differ by those phases, with stabilizers
allowed at zero rows. The induced map from the compact orbit space to the
Hausdorff momentum shell is therefore a homeomorphism.
\end{proof}

This is the known Stiefel description~\cite{HenningMelia,LM2020}.
The Stiefel manifold has real dimension $4N-4$; removing the $N$ phase
directions gives the expected momentum-space dimension $3N-4$ on the
soft-free locus. For $N\geq3$, frames with zero rows make the torus
action nonfree, so the quotient map does not define a principal
$U(1)^N$ bundle globally.

For massless momenta, define
$p_i^+=E_i+p_{z,i}$ and, on the open locus $p_i^+>0$ for every $i$,
\begin{equation}
 \rho_i=\frac{p_i^+}{M},
 \qquad
 w_i=\frac{p_{x,i}+ip_{y,i}}{\sqrt{Mp_i^+}}.
 \label{eq:lightcone-variables-main}
\end{equation}
They satisfy
\begin{equation}
 \sum_i\rho_i=1,
 \qquad
 \sum_i\abs{w_i}^2=1,
 \qquad
 \sum_i\sqrt{\rho_i}\,w_i=0.
 \label{eq:lightcone-constraints-main}
\end{equation}
Writing $r^{\mathrm{LC}}=(\sqrt{\rho_i})_{i=1}^N$ and
$w=(w_i)_{i=1}^N$, these constraints give
$r^{\mathrm{LC}}\in S^{N-1}\cap\R_{>0}^N$ and
\[
 w\in\left\{z\in\mathbb C^N:
 (r^{\mathrm{LC}})^\dagger z=0,\ \norm{z}_2=1\right\}
 \cong S^{2N-3}.
\]
As $r^{\mathrm{LC}}$ varies, these orthogonal complements form a complex
rank-$(N-1)$ vector bundle over the open simplex. The base is contractible,
so the bundle and its unit-sphere bundle are trivial. The simplex--sphere
variables of Refs.~\cite{LM2020,CCKL2024} therefore give
a full-measure product parametrization on this locus, of the form
$\mathring\Delta^{N-1}\times S^{2N-3}$ after the little-group phases are
removed. At the compact boundary the chosen lightcone section degenerates
and part of the angular fiber is identified or collapses, as is also
implicit in the compact spherical description of
Ref.~\cite{BatsonEtAl2021}.

\begin{example}[Two-particle lightcone parametrization]
\label{ex:n2-lightcone-collapse}
For two massless particles, the regular lightcone parametrization gives
$(0,1)\times S^1$. At either endpoint the transverse radius vanishes and the
angular circle collapses to a single beam-collinear pole. Thus
\begin{equation}
 \frac{[0,1]\times S^1}
 {\{0\}\times S^1\ \text{collapsed and}\
  \{1\}\times S^1\ \text{collapsed}}
 \cong S^2.
 \label{eq:n2-cylinder-collapse}
\end{equation}
\end{example}

\subsection{Gram factors and the collider spectrahedron}
\label{app:gram-geometry}

For a future-timelike momentum, the bispinor is positive definite of
rank two and factors as
\begin{equation}
 \mathsf P_i
 =L_iL_i^\dagger
 =\sum_{I=1}^2\lambda_i^I(\lambda_i^I)^\dagger,
 \qquad
 L_i=(\lambda_i^1\ \lambda_i^2)\in\mathbb C^{2\times2}.
 \label{eq:massive-spinor-factor}
\end{equation}
The determinant records the mass,
\begin{equation}
 m_i^2=\det\mathsf P_i=\abs{\det L_i}^2.
 \label{eq:massive-spinor-determinant}
\end{equation}
The same factorization includes massless and zero-momentum blocks through
rank-deficient $L_i$.

In the center-of-momentum frame, define the normalized bispinors
\begin{equation}
 \widehat X_i
 =\frac{\mathsf P_i}{M}
 =\frac1M\left(E_iI_2+\bm p_i\cdot\bm\sigma\right).
 \label{eq:collider-pauli-map}
\end{equation}
Their eigenvalues are $(E_i\pm\norm{\bm p_i})/M$, so the physical
energy inequalities and momentum conservation become
\begin{equation}
 \widehat X_i\succeq0,\qquad
 \sum_i\widehat X_i=I_2,\qquad
 \det\widehat X_i
 =\frac{E_i^2-\norm{\bm p_i}^2}{M^2}
 =\left(\frac{m_i}{M}\right)^2.
 \label{eq:collider-psd-constraints}
\end{equation}
Consequently,
\begin{equation}
 \boxed{
 \cK_N(M)\cong
 \left\{(\widehat X_1,\ldots,\widehat X_N)\in
 \operatorname{Herm}(2)^N:
 \widehat X_i\succeq0,\ \sum_i\widehat X_i=I_2\right\}.}
 \label{eq:collider-spectrahedron}
\end{equation}
After eliminating the last block, the right-hand side is the single linear
matrix inequality
\begin{equation}
 \operatorname{diag}\left(
 \widehat X_1,\ldots,\widehat X_{N-1},
 I_2-\sum_{i<N}\widehat X_i\right)\succeq0.
 \label{eq:collider-lmi}
\end{equation}
Thus $\cK_N(M)$ is a bounded spectrahedron, identified with
the convex set of labeled $N$-outcome qubit POVMs~\cite{Watrous2018}.

To obtain the corresponding spinor frame, set $A_i=L_i/\sqrt M$, so
$\widehat X_i=A_iA_i^\dagger$, and stack the adjoints:
\begin{equation}
 \mathcal A=
 \begin{pmatrix}
 A_1^\dagger\\ \vdots\\ A_N^\dagger
 \end{pmatrix}
 \in\mathbb C^{2N\times2}.
 \label{eq:full-stiefel-stack}
\end{equation}
The normalization in Eq.~\eqref{eq:collider-psd-constraints} is exactly
$\mathcal A^\dagger\mathcal A=I_2$, so
$\mathcal A\in V_2(\mathbb C^{2N})$.

Blockwise transformations $A_i\mapsto A_iU_i$, with $U_i\in U(2)$,
give all and only alternative Gram factorizations of the same tuple,
including rank-deficient blocks. Indeed, if
$AA^\dagger=BB^\dagger=X\succeq0$, polar decomposition and extension of
a partial isometry to a unitary in equal finite dimensions give
$A=X^{1/2}U$ and $B=X^{1/2}V$ with $U,V\in U(2)$.
Hence $B=A(U^\dagger V)$. Conversely, right multiplication by a unitary
leaves $AA^\dagger$ unchanged.%
\footnote{For full-rank factors, fixing the conventional determinant phase
$\det L_i=m_i>0$ restricts this unitary freedom to the physical
$SU(2)$ massive little group~\cite{Nima, KLEISS1986159}. The $U(2)$ Gram-factor
description leaves that phase unfixed and also includes the
rank-deficient massless and zero-momentum limits.}
The induced continuous bijection from the compact orbit space to the
Hausdorff spectrahedron is therefore a homeomorphism:
\begin{equation}
 \boxed{\cK_N(M)\cong V_2(\mathbb C^{2N})/U(2)^N.}
 \label{eq:full-stiefel-quotient}
\end{equation}
If $\mathcal A_i=A_i^\dagger$ denotes a row block, then
$A_i\mapsto A_iU_i$ induces
$\mathcal A_i\mapsto U_i^\dagger\mathcal A_i$.
Over the all-massive relative interior the action is free, giving
\begin{equation}
 \dim_{\R}V_2(\mathbb C^{2N})-\dim_{\R}U(2)^N
 =(8N-4)-4N=4N-4.
\end{equation}
At a rank-one block the stabilizer is a $U(1)$, and at a zero block it is
$U(2)$, giving the corresponding orbit-type stratification.

A fixed-mass shell is recovered by imposing
\begin{equation}
 \det\widehat X_i=\left(\frac{m_i}{M}\right)^2,
 \qquad\text{equivalently}\qquad
 \abs{\det A_i}=\frac{m_i}{M}.
 \label{eq:fixed-mass-determinant-slice}
\end{equation}
Thus
\begin{equation}
 \Phi_N(M;\bm m)
 \cong
 \left\{\mathcal A\in V_2(\mathbb C^{2N}):
 \abs{\det A_i}=\frac{m_i}{M}\ \text{for every }i\right\}/U(2)^N.
 \label{eq:fixed-mass-stiefel-level-quotient}
\end{equation}
Its topology is established in Appendix~\ref{app:radial-topology}.
The massless restriction has blocks of rank at most one, with rank one on
the soft-free locus, and reduces to the
$V_2(\mathbb C^N)/U(1)^N$ description above.
Allowing the determinants to vary gives the collider spectrahedron
$\cK_N(M)$, which is homeomorphic to the collider dome
$\cC_N^+(M)$ through the coordinate map in
Eq.~\eqref{eq:collider-coordinate-homeomorphism}.

\bibliographystyle{jHEP}
\bibliography{references}
\end{document}